\documentclass[aps,pra,twocolumn,groupedaddress,showpacs]{revtex4-1}
\usepackage{graphicx}
\usepackage{amsmath}
\usepackage{nicefrac}
\begin{document}

%\openup 1em

%user-defined macros
\newcommand{\barb}{\overline{b}}
\newcommand{\barbh}{\overline{b}^{(0)}}
\newcommand{\barpsi}{\overline{\psi}}
\newcommand{\barphi}{\overline{\phi}}
\newcommand{\barbeta}{\overline{\beta}}
\newcommand{\dkap}{\frac{d\kappa}{\sqrt{2\pi}}}
\newcommand{\dkapp}{\frac{d\kappa'}{\sqrt{2\pi}}}
\newcommand{\barGamma}{\overline{\Gamma}}
\newcommand{\rhobar}{\overline{\rho}}

\preprint{pra}

\title{Strongly driven nonlinear quantum optics in microring resonators}

\author{Z. Vernon}
\email{zachary.vernon@utoronto.ca}
\author{J.E. Sipe}

\affiliation{Department of Physics, University of Toronto, 60 St. George Street, Toronto, Ontario, Canada, M5S 1A7}

\date{\today}

\begin{abstract}
We present a detailed analysis of strongly driven spontaneous four-wave mixing in a lossy integrated microring resonator side-coupled to a channel waveguide. A nonperturbative, analytic solution within the undepleted pump approximation is developed for a cw pump input of arbitrary intensity. In the strongly driven regime self- and cross-phase modulation, as well as multi-pair generation, lead to a rich variety of power-dependent effects; the results are markedly different than in the low power limit. The photon pair generation rate, single photon spectrum, and joint spectral intensity (JSI) distribution are calculated. Splitting of the generated single photon spectrum into a doublet structure associated with both pump detuning and cross-phase modulation is predicted, as well as substantial narrowing of the generated signal and idler bandwidths associated with the onset of optical parametric oscillation at intermediate powers. Both the correlated and uncorrelated contributions to the JSI are calculated, and for sufficient powers the uncorrelated part of the JSI is found to form a quadruplet structure. The pump detuning is found to play a crucial role in all of these phenomena, and a critical detuning is identified which divides the system behaviour into distinct regimes, as well as an optimal detuning strategy which preserves many of the low-power characteristics of the generated photons for arbitrary input power.
\end{abstract}

\pacs{42.65.Lm, 42.65.Wi, 42.50.Dv}

\maketitle
\section{Introduction} \label{sec:intro}
Integrated optical microresonators continue to develop as a promising platform for generating, controlling, and measuring quantum states of light \cite{Xu2008,Xia2007,Ferrera2008,Ferrera2009,Azzini2012,Azzini2012a, Preble2015, Gentry2015,Kumar2015,Jung2013}. Advances in fabricating such chip-based structures is enabling the construction of micron-scale optical ring resonators with reported quality factors Q of over one million \cite{Ferrera2009,Razzari2010}. By exploiting the nonlinear optical response of the ring medium, combined with the massive enhancement of intraring pump intensity made possible by the large Q values of the resonator, a wide variety of nonlinear optical phenomena can be realized using relatively modest input powers. Entangled photon pair generation in silicon microrings has been demonstrated using mere $\mathrm{\mu W}$ of pump power \cite{Azzini2012,Gentry2015}, and optical parametric oscillation in a silicon nitride microring has been observed using a pump power of only 50 mW \cite{Levy2010}. Arrays of coupled silicon microrings have also been investigated as a potential source of heralded single photons \cite{Davanco2012}.

Such high-Q microrings are ideal for investigations of strongly driven nonlinear optical effects. Depending on the application, these effects can be undesirable or highly sought after: multi-pair production from an entangled photon pair source contaminates the sought after energy correlation, whereas optical parametric oscillation (OPO) \emph{only} arises in the strongly driven regime. Theoretical studies of quantum nonlinear optics in integrated microresonators have typically treated the nonlinearity perturbatively \cite{Vernon2015,Chen2011,Camacho2012,Helt2010, Yang2007,Yang2007a,Helt2012,Liscidini2012}, which limits calculations to quantities relating to a single generated photon pair.

Recently we presented a general theoretical treatment of photon pair generation arising from spontaneous four-wave mixing (SFWM) in microring resonators, fully accounting for the quantum effects of scattering losses within the resonator \cite{Vernon2015}. As our focus was on the effects of such losses, we limited ourselves to the low power regime in which a perturbative solution of the relevant equations of motion provides an adequate description of the pair generation process. In this work we extend our analysis to a more strongly driven regime, where perturbative strategies are inadequate and competing nonlinear effects, including self-phase modulation (SPM) and cross-phase modulation (XPM), become important. We restrict ourselves to one of the most common pump states used in experiment, that of a coherent, narrowband continuous wave pump beam, for which a nonperturbative, analytic solution to the semiclassical equations of motion is achievable within the undepleted pump approximation. This approximation limits us to pump intensities below the onset of OPO, the threshold for which is clearly indicated by our equations; OPO in such structures will be the subject of a later communication. Even below the OPO threshold, the subtle interplay between the various nonlinear terms that couple the ring modes, as well as the effects of multiple photon pair generation, give rise to a rich variety of nonlinear optical phenomena that are accessible by varying only two input parameters, namely the pump intensity and detuning.

In Sec. \ref{sec:hamiltonian_and_fields} we begin by assembling the relevant Hamiltonian and field operators for the ring-channel system. In Sec. \ref{sec:eqns_of_motion} a brief review and summary of our earlier \cite{Vernon2015} theoretical framework is presented, wherein the system's dynamics are reduced to a set of coupled ordinary differential equations for the ring operators alone. Steady state solutions for the pump mode, incorporating the effects of SPM and scattering losses, are developed in Sec. \ref{sec:pump_dynamics}, and the stability of those solutions is studied. The equations of motion for the signal and idler modes are then solved in Sec. \ref{sec:signal_idler_dynamics}, enabling the calculation of physical quantities including the photon generation rate, single photon power spectrum, and joint spectral intensity distribution. For each of these measurable quantities the corresponding predictions at low and high pump powers are compared, and we identify a set of experimental features, or ``smoking guns," that distinguish the qualitative behaviour at high pump powers from that at low pump powers.

\section{Hamiltonian and fields}\label{sec:hamiltonian_and_fields}
\begin{figure}
\includegraphics[width=1.0\columnwidth]{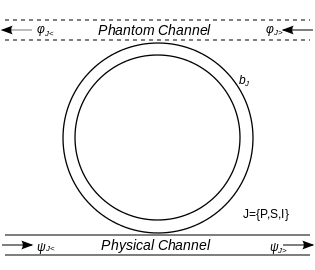}
\caption{Integrated ring-channel system geometry with labelled ring modes and incoming and outgoing outgoing channel fields. Photons generated in the ring may exit to the physical channel or be lost to the upper effective ``phantom channel".} \label{fig:ring_diagram}
\end{figure}

We consider an integrated microring resonator side-coupled to a channel waveguide, as illustrated in Fig. \ref{fig:ring_diagram}. We assume the ring size and quality factor Q have been chosen such that the ring accommodates individual resonant modes which are well separated in frequency; that is, we are in the high finesse limit. While a simple generalization of our framework can be used to treat arbitrarily many ring resonances, we restrict our model for the time being to contain only three ring modes of interest. The full system Hamiltonian can then be written \cite{Vernon2015} as
\begin{eqnarray}\label{main_hamiltonian}
H = H_{\mathrm{channel}} + H_{\mathrm{ring}} + H_{\mathrm{coupling}} + H_{\mathrm{bath}},
\end{eqnarray}
wherein $H_{\mathrm{channel}}$ refers to the channel fields, $H_{\mathrm{ring}}$ to the ring modes, $H_{\mathrm{coupling}}$ to the coupling between the channel and ring, and $H_{\mathrm{bath}}$ to any modes into which ring photons may be lost, as well the coupling of those modes to the ring modes. Introducing channel fields $\psi_J(z)$, the channel Hamiltonian is
\begin{eqnarray}\label{H_channel_defn}
H_{\mathrm{channel}} &=& \sum_J \Bigg[ \hbar\omega_J \int dz\; \psi_J^\dagger(z)\psi_J(z) \nonumber \\ 
&+& \frac{i\hbar v_J}{2}\int dz\; \left( \frac{d\psi_J^\dagger(z)}{dz}\psi_J(z) - \mathrm{H.c.}\right)\Bigg],
\end{eqnarray}
where the fields satisfy the usual commutation relations
\begin{eqnarray}\label{commutators}
\left[\psi_J(z),\psi_{J'}(z')\right] &=& 0,\nonumber \\
\left[\psi_J(z),\psi_{J'}^\dagger(z')\right] &=& \delta(z-z')\delta_{JJ'}.
\end{eqnarray}
The index $J\in\{P,S,I\}$ runs over three fields of interest, respectively labelled $P$, $S$ and $I$ for pump, signal and idler, with corresponding reference frequencies $\omega_J$ and propagation speeds $v_J$. Each field $\psi_J$ contains frequency components centred at $\omega_J$, taken to be the resonant frequency of the corresponding ring mode, and ranges over a bandwidth that does not overlap with those of other fields $\psi_{J'}$, but involving excitation over sufficiently long distances that the Dirac $\delta$ function in (\ref{commutators}) is a good approximation \cite{Vernon2015}. By allowing the fields to have different propagation speeds we include the possibility of group velocity dispersion between the different channel fields. The Hamiltonian (\ref{H_channel_defn}) does assume group velocity dispersion within each channel field is negligible, but it is straightforward to include arbitrary dispersion. The spatial co-ordinate $z$ ranges from $z=-\infty$ to $z=+\infty$ with the coupling to the ring assumed to take place at a single point $z=0$. Within this point coupling approximation the coupling Hamiltonian becomes
\begin{eqnarray}
H_{\mathrm{coupling}} = \sum_J\left(\hbar\gamma_J b_J^\dagger \psi_J(0) + \mathrm{H.c.}\right),
\end{eqnarray}
in which we have introduced ring-channel coupling coefficients $\gamma_J$, as well as discrete ring mode annihilation operators $b_J$. In addition to the physical channel, to simulate scattering losses in the ring we include an extra ``phantom channel" into which ring photons can be lost. The phantom channel similarly accommodates three fields $\phi_J(z)$ with respective propagation speeds $u_J$ and coupling coefficients $\mu_J$, and is represented as $H_\mathrm{bath}$ by a channel and coupling Hamiltonian identical to those for the physical channel \cite{Vernon2015}.

The Hamiltonian for the ring modes can be written as
\begin{eqnarray}
H_{\mathrm{ring}} = \sum_J \hbar\omega_J b_J^\dagger b_J + H_{\mathrm{NL}},
\end{eqnarray}
where $H_{\mathrm{NL}}$ includes all the nonlinearity in the system. Since the fields will be most intense within the ring resonator, we neglect channel nonlinearities and take $H_{\mathrm{NL}}$ to contain only ring mode operators. In this work we consider effects arising from the third-order nonlinear susceptibility in the ring, taking
\begin{eqnarray}
H_{\mathrm{NL}} &=& \left(\hbar\Lambda b_P b_P b_S^\dagger b_I^\dagger + \mathrm{H.c.}\right) + \hbar\eta b_P^\dagger b_P^\dagger b_P b_P \nonumber \\
&+& \hbar\zeta\left(b_S^\dagger b_P^\dagger b_S b_P + b_I^\dagger b_P^\dagger b_I b_P\right).
\end{eqnarray}
The first term is responsible for SFWM, in which two pump photons are converted to a signal and idler photon pair. The second leads to SPM of the pump, while the latter two are responsible for XPM between the pump and signal and idler modes. It is safe to neglect SPM and XPM terms that involve only the signal and idler modes, since the power in those modes will be small compared to that in the pump mode. While we focus in this work on SFWM involving a single pump mode, it is straightforward to incorporate multiple pump modes into our model. The nonlinear coupling coefficients $\Lambda$, $\eta$ and $\zeta$ are not independent, as they arise from the same nonlinear susceptibility, but we formally leave them arbitrary for the time being so that the effects of each term in $H_{\mathrm{NL}}$ can more easily be identified. Obtaining expressions for these constants depends on the approximations used to derive the nonlinear sector of the ring Hamiltonian. We present our derivation of this Hamiltonian and the associated constants $\Lambda$, $\eta$ and $\zeta$ in Appendix \ref{appendix:lambda}, arriving at an estimate of
\begin{eqnarray}
\Lambda \approx \frac{\hbar\omega_P^2cn_2}{n^2 V_{\mathrm{ring}}},
\end{eqnarray}
with $\eta=\Lambda/2$ and $\zeta=2\Lambda$. In this expression $n_2$ refers to the nonlinear refractive index of the ring material, $n$ to the linear refractive index, and $V_{\mathrm{ring}}$ to the volume of the ring mode. For the silicon nitride rings used in typical experiments \cite{Levy2010}, with $n_2\approx 2.4\times 10^{-19}\mathrm{m^2/W}$ \cite{Ikeda2008} this yields $\Lambda\sim 10\;\mathrm{Hz}$. For typical silicon rings \cite{Azzini2012}, with $n_2 \approx 2.7\times 10^{-18}\mathrm{m^2/W}$ \cite{Boyd2008} this calculation predicts $\Lambda \sim 10^3\;\mathrm{Hz}$.

\section{Equations of motion}\label{sec:eqns_of_motion}
The Heisenberg equations of motion for the field operators $\psi_J(z,t)$ and $\phi_J(z,t)$ and the ring operators $b_J(t)$ follow from the Hamiltonian (\ref{main_hamiltonian}), and can be simplified by the introduction of auxiliary quantities\cite{Vernon2015}; here we summarize the results.

The equations of motion for the channel fields are
\begin{eqnarray}\label{basic_field_eqn}
\left(\frac{\partial}{\partial t} + v_J\frac{\partial}{\partial z} + i\omega_J\right)\psi_J(z,t) = -i\gamma_J b_J(t) \delta(z),
\end{eqnarray}
with similar expressions obeyed by the phantom channel fields $\phi_J(z,t)$. Note that the solutions to these equations contain a discontinuity at $z=0$, which is a consequence of our point-coupling assumption. To avoid explicitly dealing with this discontinuity, it is helpful to introduce formal channel fields which we identify as those fields which are incoming and outgoing with respect to the coupling point. We define the incoming field $\psi_{J<}(z,t)$ by
\begin{eqnarray}
\psi_{J<}(z,t) = \psi_J(z,t)\;\;\mathrm{for}\; z<0,
\end{eqnarray}
and extend it to $z\geq 0$ by requiring everywhere that it satisfy the homogeneous version of (\ref{basic_field_eqn}),
\begin{eqnarray}
\left(\frac{\partial}{\partial t} + v_J\frac{\partial}{\partial z} + i\omega_J\right)\psi_{J<}(z,t) = 0.
\end{eqnarray}
This confers a false future on $\psi_{J<}(z,t)$, corresponding to the free evolution of the incoming field without any coupling to the ring. We similarly define the outgoing field $\psi_{J>}(z,t)$ by taking
\begin{eqnarray}
\psi_{J>}(0,t) = \psi_J(z,t)\;\;\mathrm{for}\; z>0, 
\end{eqnarray}
and demanding that for all $z$
\begin{eqnarray}
\left(\frac{\partial}{\partial t} + v_J\frac{\partial}{\partial z} + i\omega_J\right)\psi_{J>}(z,t) = 0,
\end{eqnarray}
giving $\psi_{J>}(z,t)$ a false past to the left of the coupling point. By an identical procedure we may define the incoming and outgoing phantom channel fields $\phi_{J<}(z,t)$ and $\phi_{J>}(z,t)$. Since we will primarily be concerned with the properties of the photons generated in the ring, which exit to one of the channels and propagate to positive $z$, all calculations involving the ring's output will be carried out on the outgoing fields $\psi_{J>}(z,t)$. Our goal is therefore to construct an explicit solution for these fields in terms of the incoming fields $\psi_{J<}(z,t)$. Indeed, since these fields freely propagate, the field at large positive $z$ (where any measurements on the generated photons would occur) is entirely determined by the outgoing field at $z=0$,
\begin{eqnarray}
\psi_J(z,t) = e^{-i\omega_J z/v_J}\psi_{J>}(0,t-\nicefrac{z}{v_J})\;\;\mathrm{for}\;z>0.
\end{eqnarray}
It therefore suffices to construct a solution for $\psi_{J>}(0,t)$, which can be very simply related to the incoming field $\psi_{J<}(0,t)$ and the corresponding ring operator $b_J(t)$ \cite{Vernon2015} via
\begin{eqnarray}\label{channel_transformation}
\psi_{J>}(0,t) = \psi_{J<}(0,t) - \frac{i\gamma_J}{v_J}b_J(t).
\end{eqnarray}

For each operator $\mathcal{O}_J(t)$ it will be convenient to define the corresponding slowly-varying barred operator $\overline{\mathcal{O}}_J(t)$,
\begin{eqnarray}
\overline{\mathcal{O}}_J(t) = e^{i\omega_J t}\mathcal{O}_J(t).
\end{eqnarray}
In terms of these quantities and the incoming and outgoing fields, the equations for the ring mode annihilation operators $\barb_J(t)$ are found to satisfy
\begin{widetext}
\begin{subequations}\label{ring_eqns_master}
\begin{eqnarray}
\left(\frac{d}{dt} + \barGamma_P + 2i\eta\barb_P^\dagger(t)\barb_P(t)\right)\barb_P(t) &=& -i\gamma_P^*\barpsi_{P<}(0,t) - i\mu_{P}^*\barphi_{P<}(0,t) - 2i\Lambda^*\barb_P^\dagger(t)\barb_S(t)\barb_I(t)e^{-i\Delta_\mathrm{ring} t},  \label{ring_pump_master} \\
\left(\frac{d}{dt} + \barGamma_S + i\zeta\barb_P^\dagger(t)\barb_P(t)\right)\barb_S(t) &=& -i\gamma_S^*\barpsi_{S<}(0,t) - i\mu_S^*\barphi_{S<}(0,t) -i\Lambda\barb_P(t)\barb_P(t)\barb_I^\dagger(t)e^{i\Delta_\mathrm{ring} t}, \label{ring_signal_master}\\
\left(\frac{d}{dt} + \barGamma_I + i\zeta\barb_P^\dagger(t)\barb_P(t)\right)\barb_I(t) &=& -i\gamma_I^*\barpsi_{I<}(0,t) - i\mu_I^*\barphi_{I<}(0,t) -i\Lambda\barb_P(t)\barb_P(t)\barb_S^\dagger(t)e^{i\Delta_\mathrm{ring} t}, \label{ring_idler_master}
\end{eqnarray}
\end{subequations}
\end{widetext}
where we have introduced the ring mode detuning
\begin{eqnarray}
\Delta_\mathrm{ring} = \omega_S + \omega_I - 2\omega_P,
\end{eqnarray}
as well as the total effective linewidths $\barGamma_J$,
\begin{eqnarray}
\barGamma_J=\Gamma_J + M_J,
\end{eqnarray}
where $\Gamma_J$ and $M_J$ denote the damping rates associated with the physical channel and phantom channel couplings, respectively:
\begin{eqnarray}
\Gamma_J &=& \frac{|\gamma_J|^2}{2v_J}, \nonumber \\
M_J &=& \frac{|\mu_J|^2}{2u_J}.
\end{eqnarray}
These total damping rates can be simply related to the quality factors $Q_J$ of the resonator modes; for example, for the pump resonance
\begin{eqnarray}
Q_P=\frac{\omega_P}{\barGamma_P},
\end{eqnarray}
which yields $Q_P\sim 10^6$ for a ring with $\barGamma_P=1$ GHz given a pump with wavelength $\lambda=1550$ nm. The coupled set of driven, damped ordinary differential equations (\ref{ring_eqns_master}) fully describes the nonlinear dynamics of the ring-channel system. Combined with the channel transformation (\ref{channel_transformation}), a solution to this system of equations permits the calculation of any measurable quantities on the outgoing photons in the channel.

It is important to note at this stage that our treatment neglects the effect of ring heating due to the large circulating pump power present in the ring. Such thermal effects are routinely observed in experimental investigations of microring systems, and typically manifest as an effective power-dependent drift in the resonant frequencies of the ring as it undergoes thermal expansion \cite{Levy2010}. For slowly varying and cw pumps a simple way to account for this is through the addition of a pump photon number-dependent correction to each resonance. Our model already incorporates a similar effect: SPM and XPM of each mode are represented by precisely such terms. The inclusion of thermal resonance drift can therefore be modelled by altering the coefficients $\eta$ and $\zeta$ in Eqs. (\ref{ring_eqns_master}), which would be replaced by effective constants $\eta_{\mathrm{eff}}$ and $\zeta_\mathrm{eff}$,
\begin{eqnarray}\label{thermal_substitution}
\eta_\mathrm{eff} = \eta + \eta_\mathrm{thermal}, \nonumber \\
\zeta_\mathrm{eff} = \zeta + \zeta_\mathrm{thermal}.
\end{eqnarray}
While $\eta$ and $\zeta$ are both positive, $\eta_\mathrm{thermal}$ and $\zeta_\mathrm{thermal}$ would be negative, since as the ring expands the resonant frequencies are typically lowered \cite{Almeida2004}. Depending on the relative magnitude of the thermal drift coefficients compared to the SPM and XPM strengths, in some circumstances $\eta_\mathrm{eff}$ and $\zeta_\mathrm{eff}$ may become negative. While for the remainder of this work we neglect thermal drift of the ring resonances, so that $\eta_\mathrm{thermal}=\zeta_\mathrm{thermal}=0$, we emphasize that our conclusions do not depend sensitively on this assumption unless otherwise stated.

\section{Steady state pump solution}\label{sec:pump_dynamics}
The set of coupled equations (\ref{ring_eqns_master}) treats both the pump and signal and idler modes quantum mechanically, retaining the operator nature of $b_J(t)$ for each $J$. While this is necessary if one wishes to fully account for the nonclassical properties of the pump mode, in typical experiments \cite{Azzini2012a, Azzini2012} the system is pumped by a coherent laser beam or pulse. In such situations the initial pump state is described by setting each incoming pump mode to a coherent state. The pump field can then be well approximated by its expectation value, which is a classical function of time. To implement this semiclassical approximation we take
\begin{eqnarray}
\barb_P(t)\rightarrow \barbeta_P(t) = \langle \barb_P(t) \rangle.
\end{eqnarray}
In addition to treating the pump classically, we also implement the undepleted pump approximation. In the equation for the ring pump mode (\ref{ring_pump_master}) the term involving $\barb_P^\dagger \barb_S \barb_I$ accounts for the effect on the pump mode when a signal-idler photon pair is produced. Neglecting such effects, we drop this term and instead take the semiclassical pump amplitude $\barbeta_P(t)$ to satisfy
\begin{eqnarray}\label{intermediate_pump_eqn}
\bigg( \frac{d}{dt} + \barGamma_P &+& 2i\eta|\barbeta_P(t)|^2 \bigg)\barbeta_P(t)\nonumber \\ &=& -i\gamma_P^*\langle\barpsi_{P<}(0,t)\rangle,
\end{eqnarray}
in which we have assumed $\langle \barphi_{P<}(0,t) \rangle = 0$, so that there is no incoming pump energy in the phantom channel. Note that while this approximation amounts to neglecting pump depletion due to photon pair generation, linear pump losses are still accounted for in our model, as evidenced by the presence of the damping term $\barGamma_P$ in Eq. (\ref{intermediate_pump_eqn}). 

In this work we consider the case of a continuous wave (cw) pump beam injected in to the channel, so that
\begin{eqnarray}
\langle \barpsi_{P<}(0,t) \rangle = \frac{p}{\gamma_P^*}e^{-i\Delta_P t},
\end{eqnarray}
where $\Delta_P$ is the detuning of the injected pump from the ring pump resonance, and $p$ is a constant related to the input pump power $P_{\mathrm{in}}$ in the channel at the coupling point via
\begin{eqnarray}
p = \sqrt{\frac{2\Gamma_P P_\mathrm{in}}{\hbar\omega_P}}.
\end{eqnarray}
In steady state, after the ring pump mode has come to equilibrium with the channels, we expect there to be a constant average number of pump photons $N_P$ in the ring, where
\begin{eqnarray}
N_P=\lim_{t\to\infty}|\barbeta_P(t)|^2.
\end{eqnarray}
Defining $\widetilde{\beta}_P(t) = e^{i\Delta_P t}\barbeta_P(t)$, from Eq. (\ref{intermediate_pump_eqn}) we have
\begin{eqnarray}\label{pump_tilde_eqn}
\left( \frac{d}{dt} + \barGamma_P + i(2\eta|\barbeta_P(t)|^2 - \Delta_P) \right)\widetilde{\beta}_P(t) = -ip.
\end{eqnarray}
It is not difficult to show that $N_P$ will be constant only when $\widetilde{\beta}_P(t)$ has both constant amplitude and constant phase, so that $d\widetilde{\beta}_P(t)/dt=0$.  Setting this time derivative to zero in the above equation and taking the modulus squared of the result, we find that in steady state $N_P$ must be a root of the cubic equation 
\begin{eqnarray}\label{pump_cubic_root_eqn}
C_P(N_P)=0,
\end{eqnarray}
where
\begin{eqnarray}\label{pump_cubic}
& &C_P(N_P) \equiv \nonumber \\ &4&\eta^2N_P^3 - 4\eta\Delta_P N_P^2 + (\barGamma_P^2 + \Delta_P^2)N_P - |p|^2.
\end{eqnarray}
In the absence of SPM (when $\eta \to 0$, or when the input power is very small), $N_P$ is related to the incoming power by a simple linear function,
\begin{eqnarray}
N_P = \frac{|p|^2}{\barGamma_P^2 + \Delta_P^2}.
\end{eqnarray}
The presence of SPM, however, complicates the task of determining $N_P$ as a function of $|p|^2$ for a given detuning $\Delta_P$ and nonlinearity $\eta$. The cubic equation (\ref{pump_cubic_root_eqn}) has in general as many as three real, positive roots. Furthermore, only some of these may correspond to \emph{stable} solutions of (\ref{intermediate_pump_eqn}). Before solving for the roots of $C_P(N_P)$, we first derive a set of criteria to assess the stability of any such solution. 

To determine whether or not a given root of (\ref{pump_cubic}) is stable, we conduct an analysis similar to that of Hoff, Nielsen and Andersen \cite{Andersen2015}. For a given constant solution $\widetilde{\beta}_P^{(0)}$ to (\ref{pump_tilde_eqn}), we define the fluctuation amplitude $\delta\beta_P(t)$ via
\begin{eqnarray}
\widetilde{\beta}_P(t) = \widetilde{\beta}_P^{(0)} + \delta\beta_P(t).
\end{eqnarray}
Keeping terms up to first order in $\delta\beta_P$, the equations of motion satisfied by $\delta\beta_P(t)$ and $\delta\beta_P^*(t)$ can be written as
\begin{eqnarray}
\frac{d}{dt}\begin{pmatrix}
\delta\beta_P(t) \\ \delta\beta_P^*(t)
\end{pmatrix} = F\begin{pmatrix}
\delta\beta_P(t) \\ \delta\beta_P^*(t)
\end{pmatrix},
\end{eqnarray}
where $F$ is the $2\times 2$ matrix given by 
\begin{eqnarray}
\lefteqn{F =}\nonumber \\ &&\begin{pmatrix}
-\barGamma_P - i(4\eta N_P - \Delta_P) & -2i\eta\left[\widetilde{\beta}_P^{(0)}\right]^2 \\
2i\eta\left[\widetilde{\beta}_P^{(0)*}\right]^2 & -\barGamma_P + i(4\eta N_P - \Delta_P)
\end{pmatrix}.
\end{eqnarray}
For a given solution to be stable, we require the real part of both eigenvalues of $F$ to be negative, so that the fluctuation term $\delta\beta_P(t)$ will decay with time. These eigenvalues are
\begin{eqnarray}
f_\pm = -\barGamma_P \pm \sqrt{4\eta^2N_P^2 - (4\eta N_P - \Delta_P)^2}.
\end{eqnarray}
Now, $\mathrm{Re}(f_-)<0$ automatically; demanding that $\mathrm{Re}(f_+)<0$ yields the condition
\begin{eqnarray}
4\eta^2 N_P^2 - (4\eta N_P - \Delta_P)^2 < \barGamma_P^2.
\end{eqnarray}
Solving this inequality, we find that any solution $N_P$ for Eq. (\ref{pump_cubic_root_eqn}) corresponds to a stable solution of Eq. (\ref{pump_tilde_eqn}) if $|\Delta_P|$ is below a ``critical detuning", $|\Delta_P| < \Delta_{\mathrm{critical}}$, where
\begin{eqnarray}
\Delta_{\mathrm{critical}} = \sqrt{3}\;\barGamma_P.
\end{eqnarray} 
When $|\Delta_P| > \Delta_{\mathrm{critical}}$, a solution $N_P$ of (\ref{pump_cubic_root_eqn}) corresponds to a stable solution of (\ref{pump_tilde_eqn}) if and only if $N_P$ lies outside a certain interval, $N_P \notin (N_-,N_+)$, where 
\begin{eqnarray}\label{stability_eqn}
N_{\pm} = \frac{1}{3\eta}\left(\Delta_P \pm \frac{1}{2}\sqrt{\Delta_P^2 - \Delta_\mathrm{critical}^2}\right).
\end{eqnarray}
\begin{figure}
\includegraphics[width=1.0\columnwidth]{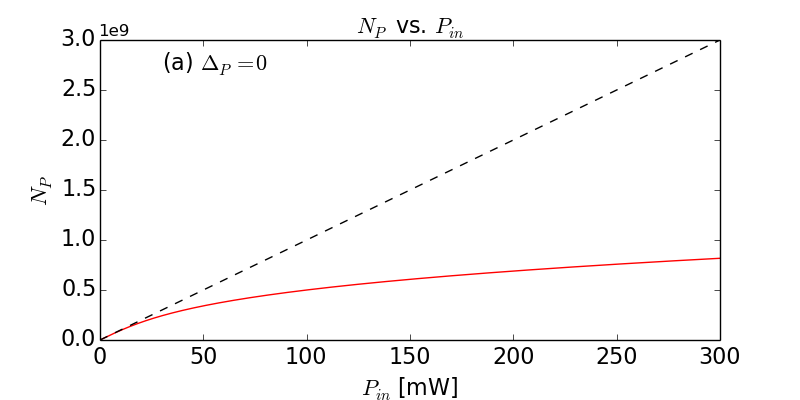}
\includegraphics[width=1.0\columnwidth]{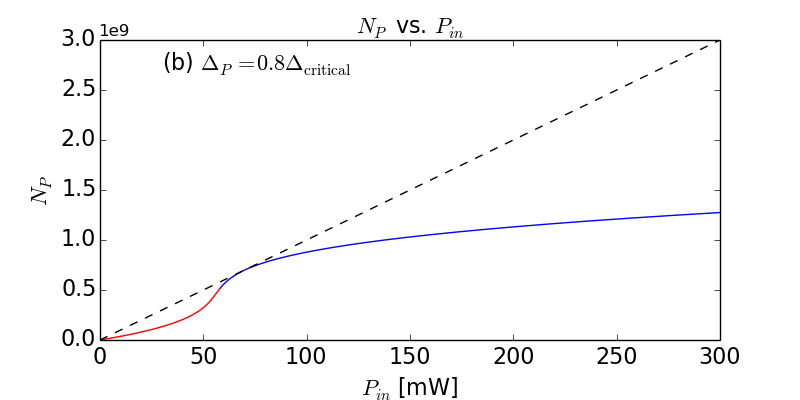}
\includegraphics[width=1.0\columnwidth]{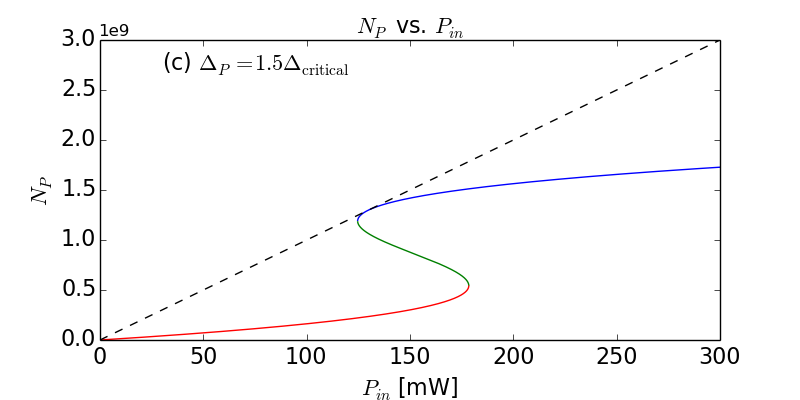}
\caption{(Colour online) Steady state average photon number in the ring pump mode as a function of channel input power with $\barGamma_P=1\;\mathrm{GHz}$, $\eta=1\;\mathrm{Hz}$ for (a) zero detuning, (b) subcritical detuning, (c) supercritical detuning. Red and blue curves indicate stable solutions, green unstable. The dashed line respresents choice of optimal detuning to maximize $N_P$ at each input power ($\Delta_P=\Delta_P^\mathrm{opt}(N_P)=2\eta N_P$).}
\label{fig:photon_number}
\end{figure}

Having established the stability criteria for a given $N_P$, we return to the task of finding real, positive roots of (\ref{pump_cubic_root_eqn}). While analytic expressions for the roots exist, it is more instructive to use indirect arguments to study their nature. Taking the derivative of $C_P$ with respect to $N_P$, we find that $dC_P/dN_P=0$ at $N_P=N_{\pm}$. Thus, for subcritical detunings ($|\Delta_P| < \Delta_\mathrm{critical})$, where the $N_{\pm}$ are not purely real, there are no local extrema -- it is easy to show that a graph of $C_P(N_P)$ is monotonically increasing and intersects the $N_P$ axis only once, leading to a single real, positive root $N_P$ which corresponds to a stable solution. On the other hand, for supercritical detunings ($|\Delta_P| > \Delta_\mathrm{critical})$, the function $C_P(N_P)$ goes through a local maximum at $N_-$ and minimum at $N_+$. The number of times $C_P(N_P)$ intersects the $N_P$ axis is then determined by the power parameter $|p|^2$; varying it translates the graph of $C_P(N_P)$ vertically. If $C_P(N_-)>0$ and $C_P(N_+)<0$, the graph of the function intersects the $N_P$ axis three times, indicating the existence of three real, positive values of $N_P$. The outer two correspond to stable solutions, while the inner root is unstable. These multiple cases are illustrated in Fig. \ref{fig:photon_number}, in which $N_P$ is plotted as a function of input power for various detunings. For values of $|\Delta_P|$ above the critical detuning of $\sqrt{3}\;\barGamma_P$ there exists a region of optical bistability, in which two stable equilibrium average pump photon numbers for a given input power are permitted, a phenomenon that has been observed experimentally in microring systems \cite{Almeida2004}. In this region the two stable solutions are separated by an unstable (and therefore physically inaccessible) range of $N_P$. Also plotted in this figure is the case of ``optimal detuning", $\Delta_P=\Delta_P^\mathrm{opt}(N_P)$, in which $\Delta_P$ is not taken to be fixed, but chosen to exactly cancel the effect of SPM as $P_\mathrm{in}$ is increased,
\begin{eqnarray}\label{optimal_detuning}
\Delta_P=\Delta_P^{\mathrm{opt}}(N_P) = 2\eta N_P,
\end{eqnarray}
which restores the simple linear relationship between $N_P$ and $|p|^2$,
\begin{eqnarray}
N_P = \frac{|p|^2}{\barGamma_P^2}.
\end{eqnarray} 
This behaviour is indicated by the dashed line in Fig. \ref{fig:photon_number}, which corresponds to a stable pump solution for all input powers, always lies on or above the fixed-detuning curves, and at each input power corresponds to the choice of detuning that maximizes $N_P$.

\section{Signal and idler dynamics}\label{sec:signal_idler_dynamics}
Having developed the steady state pump solution, we return to the signal and idler equations of motion. We first develop an exact solution to these equations, valid for a cw pump of arbitrary intensity, and then use this solution to calculate the photon pair generation rate, as well as the one- and two-photon spectra of the generated photons.
\subsection{Exact solution}
We begin by writing the equations (\ref{ring_signal_master}--\ref{ring_idler_master}) for the signal and idler ring operators in the presence of a classically described cw pump that leads to a ring pump amplitude of the form $\barbeta_P(t) = \barbeta_P e^{-i\Delta_P t}$, where $\barbeta_P$ is a constant. Letting $\widetilde{b}_x(t) = e^{i\Delta_P t}\barb_x(t)$ for $x=S,I$ we obtain
\begin{eqnarray}
\frac{d}{dt}\begin{pmatrix}
\widetilde{b}_S(t) \\ \widetilde{b}_I^\dagger(t)
\end{pmatrix} = M \begin{pmatrix}
\widetilde{b}_S(t) \\ \widetilde{b}_I^\dagger(t)
\end{pmatrix} + D(t),
\end{eqnarray}
where $M$ is the $2\times 2$ coupling matrix defined by
\begin{eqnarray}\label{M_defn}
\lefteqn{M =} \nonumber \\ 
& &\begin{pmatrix}
-\barGamma_S - i(\zeta|\barbeta_P|^2 - \Delta_P) & -i\Lambda \barbeta_P^2 \\
i\Lambda\barbeta_P^{*2} & -\barGamma_I + i(\zeta|\barbeta_P|^2 - \Delta_P)
\end{pmatrix},\;\;\;
\end{eqnarray}
and $D(t)$ the driving term responsible for quantum fluctuations from the physical and phantom channels,
\begin{eqnarray}
D(t) = \begin{pmatrix}
-ie^{i\Delta_P t}(\gamma_S^*\barpsi_{S<}(0,t) + \mu_S^*\barphi_{S<}(0,t)) \\
ie^{-i\Delta_P t}(\gamma_I\barpsi_{I<}^\dagger(0,t) + \mu_I\barphi_{I<}^\dagger(0,t)) \\
\end{pmatrix}.
\end{eqnarray}
In obtaining $M$ we have assumed the ring resonances are equally spaced, so that $\Delta_\mathrm{ring}=\omega_S + \omega_I - 2\omega_P=0$; the pump detuning, however, is left arbitrary. Previously \cite{Vernon2015} we employed a perturbative approach in the frequency domain to solve these equations, while ignoring the effects of SPM and XPM. While this provides an adequate description of the pair generation process for low pump powers, a nonperturbative strategy is needed to treat the strongly driven case. In the cw regime, where $M$ is time-independent, this coupled set of linear ordinary differential equations can be solved exactly in the time domain for arbitrary pump intensities by taking
\begin{eqnarray}
\begin{pmatrix}
\widetilde{b}_S(t) \\ \widetilde{b}_I^\dagger(t)
\end{pmatrix} = \int\limits_{-\infty}^{\;\;t} dt' G(t,t') D(t'),
\end{eqnarray}
where the $2\times 2$ matrix Green function $G(t,t')$ is given by 
\begin{eqnarray}
G(t,t') &=&	 e^{\int_{t'}^t M dt''} = e^{M\cdot(t-t')} \nonumber \\
&=&\begin{pmatrix}
g_D(t,t') & g_A(t,t') \\ g_A^*(t,t') & g_D^*(t,t')
\end{pmatrix}.
\end{eqnarray}
For simplicity we henceforth assume the ring-channel coupling constants and propagation speeds for each mode are the same, $\gamma_J=\gamma$, $v_J=u_J=v$ and $\mu_J=\mu$ so $\barGamma_J=\barGamma$ for each $J$. The matrix elements $g_D$ and $g_A$ are then given by
\begin{eqnarray}
&g_D&(t,t') = e^{-\barGamma(t-t')} \\ 
\;\;\;\;&\times& \bigg(\cosh[\rhobar(t-t')] - i\frac{\zeta N_P - \Delta_P}{\rhobar}\sinh[\rhobar(t-t')]\bigg) \nonumber
\end{eqnarray}
and
\begin{eqnarray}
g_A(t,t') = \frac{-i\Lambda\barbeta_P^2}{\rhobar}\sinh[\rhobar(t-t')],
\end{eqnarray}
in which we have introduced the dynamical parameter $\rhobar$,
\begin{eqnarray}\label{rhobar_defn}
\rhobar = \sqrt{\Lambda^2 N_P^2 - (\zeta N_P - \Delta_P)^2}.
\end{eqnarray}
Depending on the pump photon number $N_P$ and detuning $\Delta_P$, $\rhobar$ may be either purely real, purely imaginary, or exactly zero. Indeed, as will become clear in the following sections, $\rhobar$ serves as an important parameter in characterizing the system's behaviour. 

With explicit solutions written down for the ring operators $\widetilde{b}_J(t)$, we can make use of the incoming-outgoing channel field relation (\ref{channel_transformation}) to determine $\barpsi_{J>}(0,t)$. We find for the signal
\begin{eqnarray}
\lefteqn{\barpsi_{S>}(0,t) =} \\
 \int dt' &\bigg[&q_{SS}(t,t')\barpsi_{S<}(0,t') + p_{SS}(t,t')\barphi_{S<}(0,t') \nonumber \\
&+& q_{SI}(t,t')\barpsi_{I<}^\dagger(0,t') + p_{SI}(t,t')\barphi_{I<}^\dagger(0,t')\bigg], \nonumber
\end{eqnarray}
where we have introduced the temporal response functions $q_{xx'}(t,t')$ for the physical channel and $p_{xx'}(t,t')$ for the phantom channel:
\begin{eqnarray}
& &q_{SS}(t,t') = \delta(t-t') \nonumber \\
&-& \frac{|\gamma|^2}{v}\theta(t-t')e^{-(\barGamma + i\Delta_P)(t-t')} \\ 
&\times& [\cosh[\rhobar(t-t')] - i\frac{\zeta|\barbeta_P|^2 - \Delta_P}{\rhobar}\sinh[\rhobar(t-t')], \nonumber
\end{eqnarray}
and
\begin{eqnarray}
\lefteqn{q_{SI}(t,t') =} \\
& &\frac{-\gamma^2\Lambda\barbeta_P^2}{v\rhobar}\theta(t-t')e^{-i\Delta_P(t+t')}e^{-\barGamma(t-t')}\sinh[\rhobar(t-t')]. \nonumber
\end{eqnarray}
The phantom channel response functions are related to these via
\begin{eqnarray}
p_{SS}(t,t') = \frac{\mu^*}{\gamma^*}(q_{SS}(t,t') - \delta(t-t'))
\end{eqnarray}
and
\begin{eqnarray}
p_{SI}(t,t') = \frac{\mu}{\gamma}q_{SI}(t,t').
\end{eqnarray}
Similar response functions $p_{Ix}(t,t')$ and $q_{Ix}(t,t')$  can be introduced for the idler fields, which, due to our assumption of equal coupling coefficients and propagation speeds for the signal and idler fields, are identical to those for the signal: $p_{IS} = p_{SI}$, $p_{II}=p_{SS}$, $q_{IS}=q_{SI}$ and $q_{II}=q_{SS}$.

\subsection{Photon generation rate}\label{sec::pair_gen_rates}
\begin{figure}
\includegraphics[width=1.0\columnwidth]{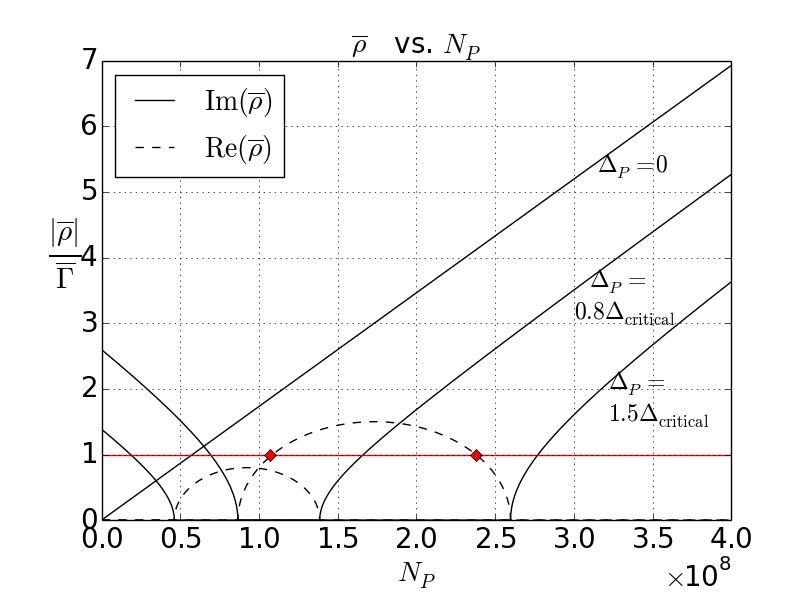}
\caption{(Colour online) Real and imaginary parts of $\rhobar$ as a function of $N_P$ for zero, subcritical and supercritical detunings, indicating that $\rhobar$ is always either purely real or purely imaginary. The red line indicates $\rhobar=\barGamma$. The transition between $\rhobar$ being purely imaginary and purely real occurs at the points $N_P=\Delta_P/3\Lambda$ and $N_P=\Delta_P/\Lambda$. For supercritical detunings, there exist points where $\rhobar=\barGamma$ (represented on the plot by red diamonds), indicating the onset of OPO behaviour. The nonlinear parameter $\Lambda$ taken as $\Lambda=10$ Hz.}\label{fig::rhobar_plot}
\end{figure}

Armed with explicit expressions for the outgoing fields $\barpsi_{S>}$ and $\barpsi_{I>}$, we can calculate any measurable quantity related to the generated signal and idler photon pairs. Of particular interest is the photon pair generation rate, one of the primary figures of merit used in assessing the practical utility of the ring-channel system. The steady state outgoing flux of signal photons $J_S$ into the physical channel can be calculated via
\begin{eqnarray}
J_S &=& \lim_{t\to\infty} v \langle \barpsi_{S>}^\dagger(0,t)\barpsi_{S>}(0,t)\rangle \nonumber \\
&=& \lim_{t\to\infty} \frac{2\barGamma}{\Gamma}\int dt' |q_{SI}(t,t')|^2.
\end{eqnarray}
Computing the integral, we find
\begin{eqnarray}
J_S = \frac{2\Gamma\Lambda^2 N_P^2}{\barGamma^2 - \rhobar^2}.
\end{eqnarray}
The nature of the scaling of $J_S$ with pump photon number $N_P$ depends intimately on the character of $\rhobar$, the behaviour of which as a function of $N_P$ for various detunings is illustrated in Fig. \ref{fig::rhobar_plot}. Recalling (\ref{rhobar_defn}), we find that $\rhobar$ is real when $N_P \in [\Delta_P/3\Lambda,\Delta_P/\Lambda]$, with $\rhobar=0$ at the endpoints of this interval, and imaginary otherwise.
\begin{figure}
\includegraphics[width=1.0\columnwidth]{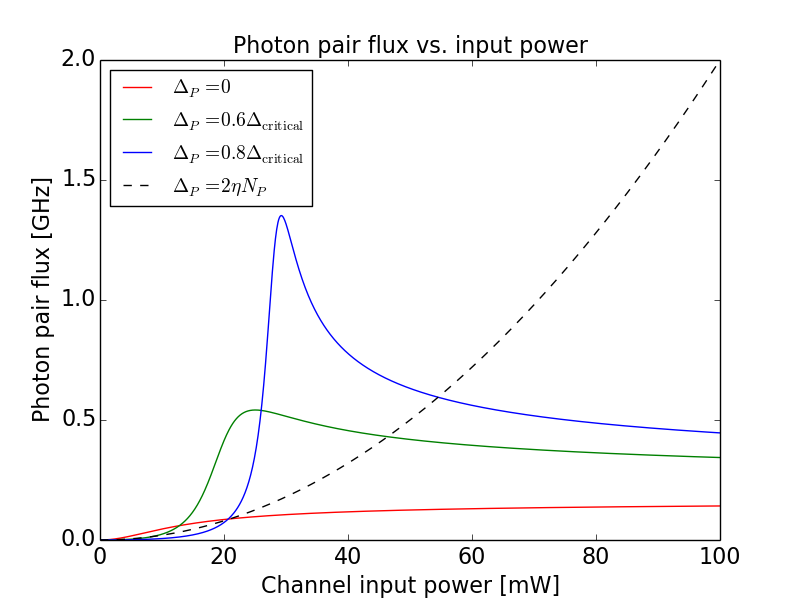}
\caption{(Colour online) Photon pair generation rate as a function of channel input power for various detunings. The dashed curve indicates the optimal detuning case. System parameters for this plot are $\eta=1\;\mathrm{Hz}$, $\barGamma=1\;\mathrm{GHz}$.}\label{fig:pair_rate_plot}
\end{figure}
For low enough $N_P$, when $\rhobar \approx i|\Delta_P|$, $J_S$ scales quadratically with the number of pump photons $N_P$. Since in this regime $N_P$ is directly proportional to the channel input power $P_{\mathrm{in}}$, the overall scaling of $J_S$ with $P_{\mathrm{in}}$ remains quadratic, in agreement with experiment \cite{Azzini2012}. As the pump power increases, however, the scaling of $J_S$ is affected by several separate power-dependent processes. 

First, for a fixed detuning $\Delta_P$, the SPM-induced drift of the pump resonance slows the scaling of $N_P$ with channel input power $P_\mathrm{in}$, as demonstrated in Fig. \ref{fig:photon_number}. Second, XPM between the pump, signal and idler modes effectively shifts the resonance lines of the signal and idler modes, compromising the resonance enhancement of the pair generation process. Finally, for supercritical detunings $|\Delta_P|>\Delta_\mathrm{critical}$, $\rhobar\to\barGamma$ when $N_P\to N_{\pm}$, where $N_{\pm}$ are the same two photon numbers that define the stability of the pump solution (\ref{stability_eqn}). In that limit the photon flux $J_S$ formally diverges. This unphysical prediction corresponds to the onset of optical parametric oscillation\cite{Levy2010,Razzari2010}. As this threshold is approached, stimulated emission leads to photon pairs being generated faster than the rate at which they are removed from the ring, preventing the system from reaching a steady state within our model. We are prevented from treating this case by our assumption of an undepleted pump. Our results are expected to be valid when the intraring conversion efficiency $E_\mathrm{ring}$ is much less than unity; this efficiency, defined as the ratio between the steady state signal (or idler) and pump photon numbers, can be expressed as
\begin{eqnarray}
E_\mathrm{ring} = \frac{J_S}{2\Gamma N_P} = \frac{\Lambda^2N_P}{\barGamma^2-\rhobar^2}.
\end{eqnarray} 
While in future work we intend to investigate the OPO regime and the associated effects of pump depletion, for the time being we restrict ourselves to regimes where $E_\mathrm{ring} \ll 1 $; in all examples presented below this inequality is satisfied. 

Perhaps most remarkable is the special regime of optimal detuning, wherein $\Delta_P$ is chosen to maximize $N_P$ at each channel input power, $\Delta_P=\Delta_P^\mathrm{opt}(N_P)$ as defined in Eq. (\ref{optimal_detuning}). For this choice $\rhobar=0$ identically for all $N_P$,
\begin{eqnarray}\label{rhobar_special}
\rhobar &=& \sqrt{\Lambda^2 N_P^2 - (\zeta N_P - \Delta_P)^2} \nonumber \\
&=& \sqrt{\Lambda^2 N_P^2 - \left(2\Lambda N_P - 2\frac{\Lambda}{2}N_P\right)^2} \nonumber \\
&=& 0.
\end{eqnarray}
The photon pair flux then maintains its quadratic scaling with both $N_P$ and channel input power $P_{\mathrm{in}}$ over its entire domain:
\begin{eqnarray}
J_S = \frac{8\Gamma^3\Lambda^2}{(\hbar\omega_P)^2\barGamma^6}P_{\mathrm{in}}^2.
\end{eqnarray}
This cancellation between the effects arising from photon pair generation, XPM, and the SPM-dependent detuning strategy $\Delta_P^\mathrm{opt}(N_P)$ arises from the simple fixed relationship between the associated nonlinear coupling strengths $\Lambda$, $\eta$ and $\zeta$. Crucial for this phenomenon is that the strength of the photon pair generation process scale quadratically with the pump photon number $N_P$. This cancellation effect would therefore not be possible using, for example, spontaneous parametric downconversion, the strength of which would scale linearly with $N_P$. The presence of thermal resonance drift would not compromise the existence of an $N_P$-dependent detuning strategy that yields $\rhobar=0$ over all $N_P$, though such a strategy would no longer correspond to that which also linearizes and maximizes the relationship between $N_P$ and channel input power.

The photon pair generation rate as a function of channel input power is plotted in Fig. \ref{fig:pair_rate_plot} for various values of $\Delta_P$ alongside this optimal detuning case. For lower powers, when SPM and XPM are negligible, a pump beam with $\Delta_P=0$ gives the best scaling of $J_S$. For intermediate powers the detuning may be tweaked to combat SPM and XPM in order to maximize $J_S$, while for high powers the optimal detuning strategy of $\Delta_P=\Delta_P^\mathrm{opt}(N_P)$ beats any fixed subcritical detuning. The behaviour of these curves suggests a simple experiment to identify the presence of nonperturbative, strongly driven effects: one could simply measure the outgoing signal or idler power as a function of pump input power for a set of fixed, subcritical pump detunings. For fixed nonzero detunings $\Delta_P<\Delta_\mathrm{critical}$, strongly driven effects are indicated by the presence of a global maximum of generated signal power at intermediate pump input power, followed by decreasing signal power approaching an asymptotic value of
\begin{eqnarray}
\hbar\omega_S\lim_{N_P\to\infty} J_S(N_P) = \hbar\omega_S\frac{2\Gamma}{3}.
\end{eqnarray}
For critically coupled ring systems $\Gamma\approx\barGamma/2$ \cite{Vernon2015}, so the asymptotic signal power can be related to the total effective ring linewidth $\barGamma$ as simply $\hbar\omega_S\barGamma/3$. If thermal detuning of the ring resonances is included this asymptotic power will be different; however, the qualitative behaviour of the signal power as a function of pump power will be unchanged.

\subsection{Single photon spectrum}
Another physical quantity of interest is the spectral lineshape of signal and idler photons that are emitted from the ring. For low power cw pumps these single photon spectra typically exhibit a Lorentzian lineshape \cite{Azzini2012,Azzini2012a,Vernon2015} with a characteristic width determined by the total effective linewidths of the microring cavity resonances. As we now demonstrate, these spectral characteristics are significantly different in the strongly pumped regime. We develop results for the signal field spectrum; the idler field will have identical properties.

\begin{figure}
\includegraphics[width=1.0\columnwidth]{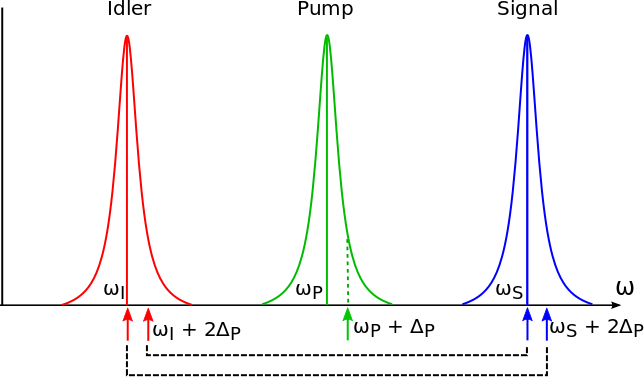}
\caption{(Colour online) Origin of splitting in the signal and idler lineshapes. Green, blue, and red curves indicate pump, signal and idler resonances, respectively, and could respresent the enhancement factor \cite{Heebner2008} that would characterize the ratio of the intensity in the ring to the incident channel intensity in a linear experiment. The dashed green line indicates a pump detuned by $\Delta_P$, which leads to a generated pair having either its signal or idler photon detuned by $\sim 2\Delta_P$ from the corresponding resonance, as indicated by the signal and idler pairs connected by dashed black lines. The presence of pairs from both cases leads to a doublet structure for both the signal and idler lineshapes.}\label{fig::splitting}
\end{figure}

We define the power spectrum \cite{Mandel1995} for the signal channel field as
\begin{eqnarray}\label{nu_S_defn1}
\lefteqn{\nu_S(\omega_s; t) =} \\
& & \lim_{T\to\infty}\frac{1}{T}\int\displaylimits_{t-T/2}^{t+T/2}  dt \int \frac{d\tau}{\sqrt{2\pi}} g^{(1)}(t,t+\tau) e^{i\omega_s \tau} \nonumber
\end{eqnarray}
where the first-order temporal coherence function $g^{(1)}(t_1,t_2)$ is defined by
\begin{eqnarray} \label{g1_defn}
g^{(1)}(t_1,t_2) &=& v \langle \barpsi_{S>}^\dagger(0,t_1)\barpsi_{S>}(0,t_2)\rangle \nonumber \\
&=& \frac{\barGamma}{\Gamma}\int dt' q_{SI}^*(t_1,t')q_{SI}(t_2,t').
\end{eqnarray}
In writing (\ref{nu_S_defn1}) we have introduced the relative frequency co-ordinate $\omega_s$, which corresponds to a frequency offset from the ring reference $\omega_S$. The physical frequency $\overline{\omega}_s$ associated with $\omega_s$ is therefore
\begin{eqnarray}
\overline{\omega}_s = \omega_S + \omega_s.
\end{eqnarray}
In the remaining sections we adopt this notation of lowercase subscripts for frequency offsets: $\omega_s$ for the signal, $\omega_p$ for the pump and $\omega_i$ for the idler.
\begin{figure*}
\includegraphics[width=1.0\columnwidth]{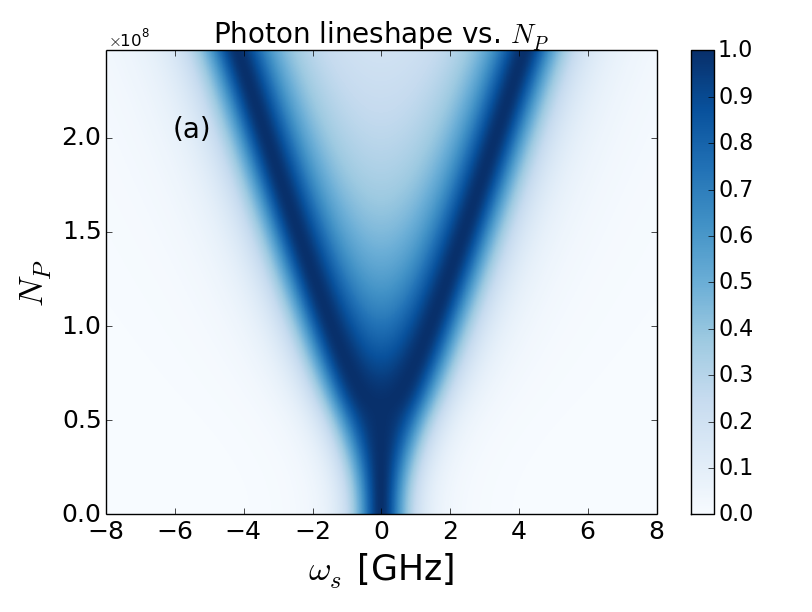}
\includegraphics[width=1.0\columnwidth]{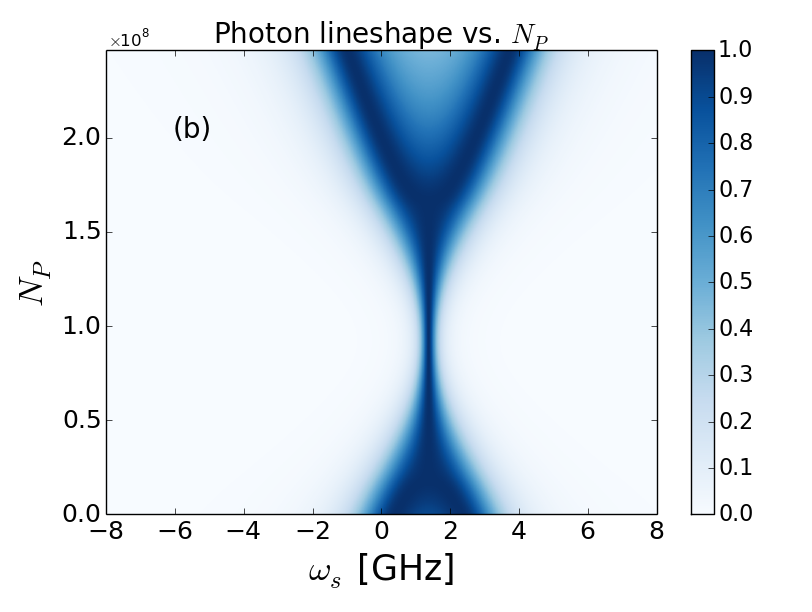}
\includegraphics[width=1.0\columnwidth]{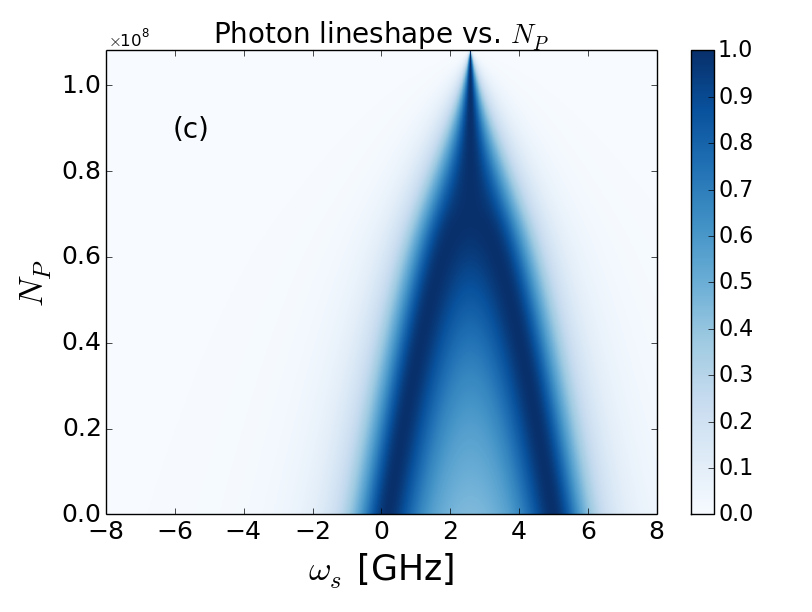}
\includegraphics[width=1.0\columnwidth]{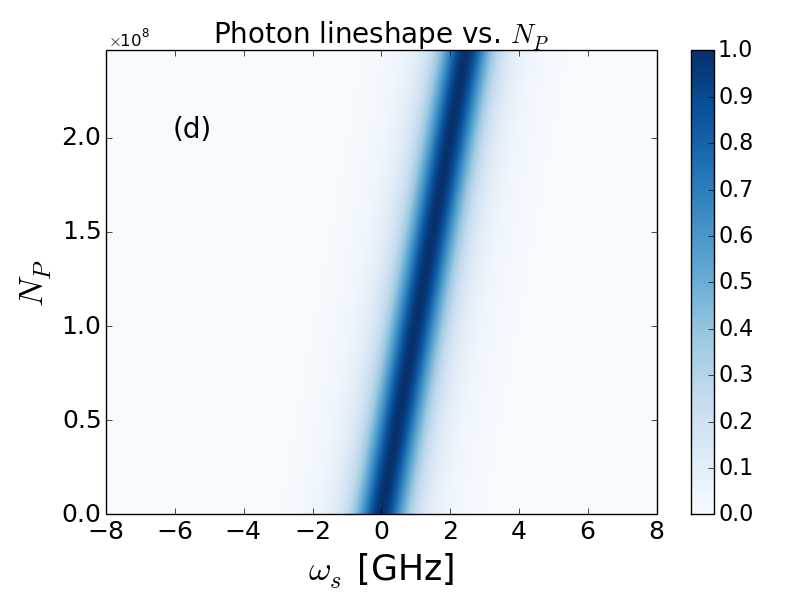}
\caption{(Colour online) Spectral lineshape $\nu_S(\omega_s)$ scaled to unit maximum vs. ring pump photon number $N_P$. For each plot we take $\barGamma=1$ GHz and $\Lambda=10$ Hz. (a) $\Delta_P=0$, (b) $\Delta_P=0.8\Delta_\mathrm{critical}\approx 1.4$ GHz, (c) $\Delta_P=1.5\Delta_\mathrm{critical}\approx2.6$ GHz, and (d) $\Delta_P=\Delta_P^\mathrm{opt}(N_P)=2\eta N_P$. The origin of the frequency axis corresponds to the ring resonance at $\omega_S$. }\label{fig::lineshapes}
\end{figure*}

Evaluating (\ref{g1_defn}) and setting $t_1=t$, $t_2=t+\tau$ we obtain
\begin{eqnarray}
\lefteqn{g^{(1)}(t,t+\tau) =}\\
& & \frac{\Gamma\Lambda^2 N_P^2}{\rhobar}e^{i\Delta\tau}e^{-\barGamma |\tau|}\frac{\rhobar\cosh[\rhobar|\tau|] + \barGamma\sinh[\rhobar|\tau|]}{\barGamma^2 - \rhobar^2}\nonumber,
\end{eqnarray}
which is independent of $t$, depending only on the relative time difference $\tau$, as would be expected for a cw pump. Taking the Fourier transform, we arrive at an expression for the lineshape,
\begin{eqnarray}\label{nu_S_defn}
\lefteqn{\nu_S(\omega_s)=} \\ 
& & \frac{4\Gamma\barGamma \Lambda^2 N_P^2}{\sqrt{2\pi}|\barGamma - \rhobar +i(\omega_s - \Delta_P)|^2|\barGamma+\rhobar+i(\omega_s-\Delta_P)|^2} \nonumber,
\end{eqnarray}
with an identical equation for the idler lineshape $\nu_I(\omega_i)$. This expression takes the form of a product of two Lorentzians. We consider first subcritical detunings. When $\rhobar$ is imaginary, these Lorentzians have identical characteristic widths $\delta\omega=\barGamma$ and are centred on $\omega_s=\Delta_P\pm|\rhobar|$. For low powers, when $\rhobar \approx i|\Delta_P|$ the spectrum is therefore peaked at $\omega_s=0$ and $\omega_s=2\Delta_P$, in agreement with the perturbative calculation \cite{Vernon2015}. This splitting is easily understood as a consequence of the tradeoff between energy conservation and resonance enhancement of the pair generation process. As illustrated in Fig. \ref{fig::splitting}, when a photon pair is produced with a detuned pump, either the signal photon \emph{or} idler photon in a pair, but not both, can be generated within a ring resonance; energy conservation then requires the other to be generated with a frequency that lies away from its corresponding resonance. This is seen in the $N_P\to 0$ limit of Fig. \ref{fig::lineshapes}(b). At sufficient pump photon number $N_P$ a similar splitting can arise from the effective XPM-induced detuning of the signal and idler ring resonances even for a pump with $\Delta_P=0$, as seen in Fig. \ref{fig::lineshapes}(a) for large $N_P$. When $\Delta_P=0$ the lineshape begins as a singly-peaked Lorentzian, eventually splitting to a doublet structure when $\rhobar$ becomes imaginary as a consequence of XPM. 

For nonzero $\Delta_P$, as $N_P$ increases, XPM effectively counters the pump detuning and the extent of this splitting is reduced as $|\rhobar|$ decreases, eventually vanishing when $N_P=\Delta_P/3\Lambda$. If $N_P$ is increased further, $\rhobar$ becomes real and ceases to contribute to spectral splitting, resulting instead in an effective correction to the linewidth. The lineshape then takes the form of a product of two Lorentzians, both centred on $\omega_s=\Delta_P$, with respective widths $\delta\omega_\pm = \barGamma \pm \rhobar$. As $\rhobar$ becomes comparable to $\barGamma$ the smaller of these two widths becomes dominant, leading to a lineshape with overall effective width $\delta\omega\approx\barGamma-\rhobar$. For subcritical detunings, as demonstrated at large $N_P$ in Fig. \ref{fig::lineshapes}(b), the spectral splitting is then resumed as $\rhobar$ once again becomes imaginary. \

For supercritical detunings, as the threshold for optical parametric oscillation is approached $\rhobar\to\barGamma$ and the bandwidth of the emitted signal and idler photons becomes arbitrarily narrow, as seen in Fig. \ref{fig::lineshapes}(c). This follows from our idealization of the pump as an indefinitely coherent cw beam; in actual experiments the bandwidth of the generated photons will become comparable to that of the pump, a phenomenon that has been observed in strongly pumped experiments on silicon nitride microrings \cite{Levy2010}. 

Finally, as shown in Fig. \ref{fig::lineshapes}(d), in the special case of optimal detuning when $\Delta_P=\Delta_P^\mathrm{opt}(N_P)$, so that $\rhobar=0$, the lineshape remains peaked at a single $N_P$-dependent frequency for each $N_P$, with unchanging characteristic width $\delta\omega=\barGamma$, precisely mimicking the low-power result at zero detuning.

Experimentally, measuring the signal or idler lineshape as a function of input power for a nonzero, subcritical detuning as in Fig. \ref{fig::lineshapes}(b) would reveal the richness of the strongly driven regime, and illustrate the behaviour of the $\rhobar$ paramater, which incorporates the effects of both XPM and pair generation.

\subsection{Joint spectral intensity}
To assess the degree of spectral correlation between the signal and idler modes, it is instructive to study the joint spectral intensity distribution of the generated photon pairs. While it is straightforward to define this quantity for a system driven by a train of weak pump pulses, in which multi-pair generation can be neglected, it is a more subtle task to craft a sensible measure of spectral correlation in the strongly driven cw regime. In particular, there is no single function that characterizes a joint probability amplitude of signal and idler photons, since in general there will be far more than two photons in the quantum state of the signal and idler modes. Furthermore, even for weak cw pumps, if one introduces outgoing channel annihilation operators $c_J(\omega_j)$ via
\begin{eqnarray}\label{amplitude_defn}
\barpsi_{J>}(0,t) = \int \frac{d\omega_j}{\sqrt{2\pi}}c_J(\omega_j)e^{-i\omega_j t}
\end{eqnarray} 
and naively calculates expectation values of the form 
$\langle c_S^\dagger(\omega_s)c_I^\dagger(\omega_i)c_S(\omega_s)c_I(\omega_i)\rangle$, the idealization of a zero-bandwidth cw pump leads to ill-defined expressions involving the square of Dirac $\delta$ distributions. 

To resolve these difficulties, in Appendix \ref{appendix:JSI} we develop a model of a typical experiment used to characterize the JSI for weakly driven systems, in which the coincidence count rate of signal and idler photons at respective frequencies $\omega_s$ and $\omega_i$ is measured. We the extend the definition of the JSI to strongly driven systems by defining the JSI to equal the calculated outcome of such an experiment for arbitrary input power. This definition reduces to the usual result for single-pair output states, and serves as a sensible measure of spectral correlation between the signal and idler fields. This coincidence rate can be written as
\begin{eqnarray}\label{JSI_definition}
I_\mathrm{corr}(\omega_s,\omega_i) &=& \frac{v^2\delta t}{(2\pi)^2}\int dt_1...\int dt_4 \\
&\bigg[& e^{i\omega_s(t_3-t_1)}e^{i\omega_i(t_4-t_2)}T(t_1)T(t_2)T(t_3)T(t_4)\nonumber \\
&\times&\langle \barpsi_{S>}^\dagger(t_1)\barpsi_{I>}^\dagger(t_2)\barpsi_{S>}(t_3)\barpsi_{I>}(t_4)\rangle\bigg],\nonumber
\end{eqnarray}
where $T(t)$ is the Fourier transform of a transmission function $\hat{T}(\omega)$ that resolves the frequencies of the signal and idler photons prior to detection, 
\begin{eqnarray}\label{T_defn}
T(t) = \int \frac{d\omega}{\sqrt{2\pi}}\hat{T}(\omega)e^{-i\omega t},
\end{eqnarray}
and $\delta t$ is the temporal resolution of the coincidence counter. In this expression the spatial dependence of the field operators $\barpsi_{J>}(z,t)$ has been suppressed; the signal and idler arms of the JSI measurement are assumed to occur at balanced distances from the ring-channel coupling point.

The four-time expectation value $\langle \barpsi_{S>}^\dagger(t_1)\barpsi_{I>}^\dagger(t_2)\barpsi_{S>}(t_3)\barpsi_{I>}(t_4)\rangle$ is found to naturally split into two parts,
\begin{eqnarray}
&v&^2\langle \barpsi_{S>}^\dagger(t_1)\barpsi_{I>}^\dagger(t_2)\barpsi_{S>}(t_3)\barpsi_{I>}(t_4)\rangle = \\
& &A^*(t_1,t_2)A(t_3,t_4) + g^{(1)}(t_1,t_3)g^{(1)}(t_2,t_4),\nonumber
\end{eqnarray}
where
\begin{eqnarray}
A(t_1,t_2) = \int dt'&	\bigg[&q_{SI}(t_1,t')q_{II}(t_2,t') \\
&+& p_{SI}(t_1,t')p_{II}(t_2,t')\bigg].\nonumber
\end{eqnarray}
The function $g^{(1)}$ is precisely the first-order coherence function defined in Eq. (\ref{g1_defn}) used to calculate the single photon spectrum,
\begin{eqnarray}
g^{(1)}(t_1,t_3) &=& \frac{\Gamma\Lambda^2 N_P^2}{\rhobar}e^{i\Delta_P(t_3-t_1)}e^{-\barGamma|t_3-t_1|} \\ 
&\times&\frac{\rhobar\cosh[\rhobar|t_3-t_1|] + \barGamma\sinh[\rhobar|t_3-t_1|]}{\barGamma^2-\rhobar^2}.\nonumber
\end{eqnarray}
The $A(t_1,t_2)$ term, after computing the integrals, is given by
\begin{eqnarray}
A(t_1,t_2)&=&\frac{\gamma^2\Lambda\barbeta_P^2}{2v}e^{-\barGamma|t_2-t_1|}e^{-i\Delta_P(t_1+t_2)} \\
&\times&\frac{[a_1\sinh[\rhobar|t_2-t_1|] + a_2\cosh[\rhobar|t_2-t_1|}{\barGamma^2-\rhobar^2},\nonumber
\end{eqnarray}
where the constants $a_1$ and $a_2$ are defined by
\begin{eqnarray}
a_1 &=& \rhobar - i\frac{\zeta N_P -\Delta_P}{\rhobar}\barGamma, \nonumber \\
a_2 &=& \barGamma -i(\zeta N_P - \Delta_P).
\end{eqnarray}
The JSI can therefore be expressed as the sum of correlated and uncorrelated terms,
\begin{eqnarray}
I(\omega_s,\omega_i) = I_{\mathrm{corr}}(\omega_s,\omega_i) + I_\mathrm{uncorr}(\omega_s,\omega_i),
\end{eqnarray}
where 
\begin{eqnarray}\label{phi_corr_defn}
\lefteqn{I_{\mathrm{corr}} (\omega_s,\omega_i) =} \\
& & \frac{\delta t}{(2\pi)^2}\bigg\vert\int d\nu_1 \int d\nu_2\hat{A}(\nu_1,\nu_2) \hat{T}(\omega_s-\nu_1)\hat{T}(\omega_i-\nu_2)\bigg \vert^2 \nonumber 
\end{eqnarray}
and
\begin{eqnarray}\label{phi_uncorr_defn}
\lefteqn{I_\mathrm{uncorr}(\omega_s,\omega_i) =\frac{\delta t}{(2\pi)^2}} \\
&\times&\int d\nu_1 \int d\nu_2\left[\hat{g}^{(1)}(\nu_1,-\nu_2)\hat{T}(\omega_s-\nu_1)\hat{T}(\omega_s - \nu_2)\right] \nonumber \\
&\times& \int d\nu_1' \int d\nu_2'\left[\hat{g}^{(1)}(\nu_1',-\nu_2')\hat{T}(\omega_i-\nu_1')\hat{T}(\omega_i - \nu_2')\right]. \nonumber
\end{eqnarray}
As indicated by their labels, $I_\mathrm{uncorr}$ can be expressed as a separable product of functions of $\omega_s$ and $\omega_i$, while $I_\mathrm{corr}$ cannot. Each is expressed as a convolution of the Fourier transforms $\hat{A}(\nu_1,\nu_2)$ and $\hat{g}^{(1)}(\nu_1,\nu_2)$ of the $A(t_1,t_2)$ and $g^{(1)}(t_1,t_2)$ functions,
\begin{eqnarray}
\hat{A}(t_1,t_2)=\int \frac{dt_1}{\sqrt{2\pi}} \int \frac{dt_2}{\sqrt{2\pi}}A(t_1,t_2)e^{i\nu_1t_1}e^{i\nu_2t_2},
\end{eqnarray}
and similarly for $g^{(1)}(\nu_1,\nu_2)$, with the transmission filter function $\hat{T}(\nu_1)\hat{T}(\nu_2)$. The $A$ and $g^{(1)}$ functions are determined by the dynamics of the signal and idler modes in the ring, while their convolution with the $T$ functions reflects the frequency averaging that arises from the finite resolution of a realistic JSI measurement.

Computing the Fourier transform, we find for $\hat{A}$
\begin{widetext}
\begin{eqnarray}
\hat{A}(\nu_1,\nu_2) = \Gamma^2\Lambda^2 N_P^2 \delta(\nu_1+\nu_2-2\Delta_P)\left(\frac{1-i\frac{\zeta N_P-\Delta_P}{\rhobar}}{(i\Delta\nu - \barGamma + \rhobar)(-i\Delta\nu-\barGamma+\rhobar)} + \frac{1+i\frac{\zeta N_P-\Delta_P}{\rhobar}}{(i\Delta\nu - \barGamma - \rhobar)(-i\Delta\nu-\barGamma-\rhobar)}\right),
\end{eqnarray}
with $\Delta\nu = (\nu_1 - \nu_2)/2$. The term in parentheses multiplying the $\delta$ function varies on the scale of $\barGamma$. Assuming that the measurement frequency resolution $\delta\omega_\mathrm{trans}$ is much narrower than this, the slowly varying term can be pulled out of the integrals in (\ref{phi_corr_defn}), leaving
\begin{eqnarray}\label{phi_corr_expression}
I_\mathrm{corr}(\omega_s,\omega_i) &\approx& \delta t\Gamma^2\Lambda^2N_P^2 [D(\omega_s-\Delta_P,\omega_i-\Delta_P)]^2 \\ 
&\times& \bigg\vert\frac{1-i\frac{\zeta N_P-\Delta_P}{\rhobar}}{(i(\omega_i-\Delta_P) - \barGamma + \rhobar)(-i(\omega_i-\Delta_P)-\barGamma+\rhobar)} + \frac{1+i\frac{\zeta N_P-\Delta_P}{\rhobar}}{(i(\omega_i-\Delta_P) - \barGamma - \rhobar)(-i(\omega_i-\Delta_P)-\barGamma-\rhobar)}\bigg\vert^2 \nonumber
\end{eqnarray}
\end{widetext}
where
\begin{eqnarray}
\lefteqn{D(\omega_s,\omega_i) =} \\
& &\frac{1}{2\pi}\int d\nu_1 \int d\nu_2 \delta(\nu_1+\nu_2)\hat{T}(\omega_s-\nu_1)\hat{T}(\omega_i-\nu_2).\nonumber
\end{eqnarray}
The function $D(\omega_s,\omega_i)$ can be interpreted as the ``smoothed" version of the Dirac $\delta(\omega_s + \omega_i)$ distribution, and arises from the finite bandwidth of the JSI measurement scheme; $D(\omega_s-\Delta_P,\omega_i-\Delta_P)$ is sharply peaked and uniform along the energy-conserving antidiagonal line $\omega_s+\omega_i-2\Delta_P=0$ with characteristic width $\delta\omega_\mathrm{trans}$ (the measurement resolution) in the direction orthogonal to that line.

Finally, taking the Fourier transform of $g^{(1)}(t_1,t_3)$, we find
\begin{eqnarray}
\lefteqn{\hat{g}^{(1)}(\nu_1,-\nu_2) = \delta(\nu_1-\nu_2) }\\
&\times& \frac{4\barGamma\Gamma\Lambda^2N_P^2}{|\barGamma-\rhobar+i(\nu_1-\Delta_P)|^2|\barGamma+\rhobar+i(\nu_1-\Delta_P)|^2}.\nonumber
\end{eqnarray}
As with $\hat{A}$, apart from the $\delta$ function this is slowly varying compared to the measurement resolution; the term multiplying the $\delta$ function can be pulled out of the integral in Eq. (\ref{phi_uncorr_defn}). The uncorrelated contribution to the JSI $I_\mathrm{uncorr}$ is therefore well approximated by
\begin{eqnarray}\label{phi_uncorr_expression}
I_\mathrm{uncorr}(\omega_s,\omega_i)\approx \frac{\delta t}{2\pi}\bigg\vert \int d\omega|\hat{T}(\omega)|^2\bigg\vert^2\nu_S(\omega_s)\nu_I(\omega_i),\;\;\;
\end{eqnarray}
where $\nu_S(\omega_s)$ and $\nu_I(\omega_i)$ are precisely the single-photon lineshape functions given by Eq. (\ref{nu_S_defn}) as derived in the previous section. The uncorrelated part of the JSI is therefore proportional to the simple product the signal and idler lineshapes.

\begin{figure}
\includegraphics[width=1.0\columnwidth]{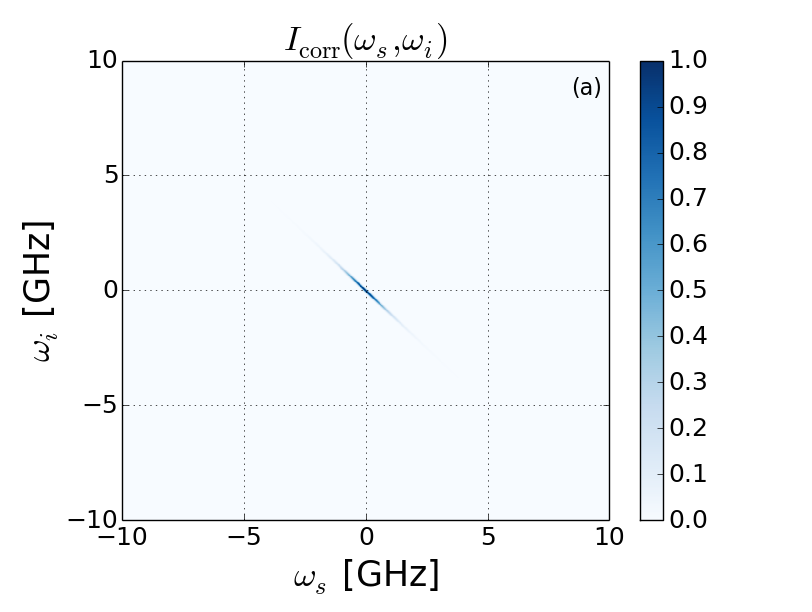}
\includegraphics[width=1.0\columnwidth]{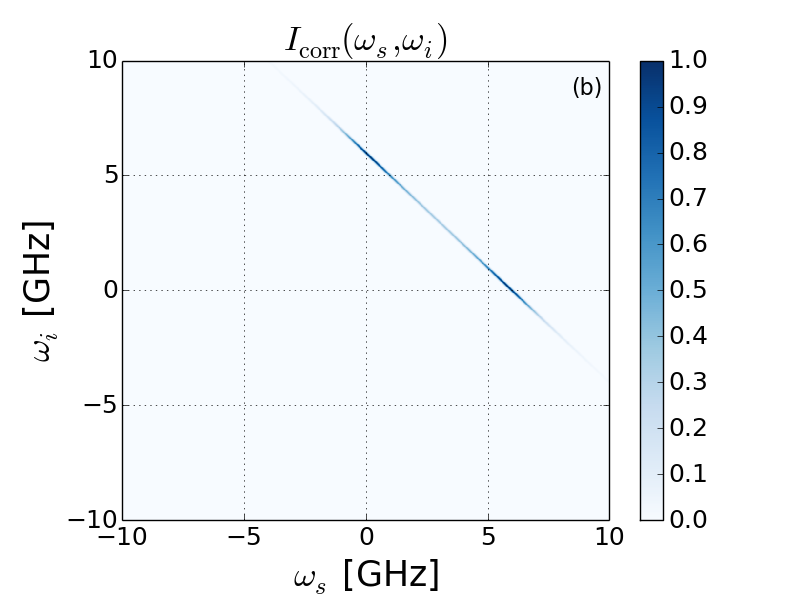}
\includegraphics[width=1.0\columnwidth]{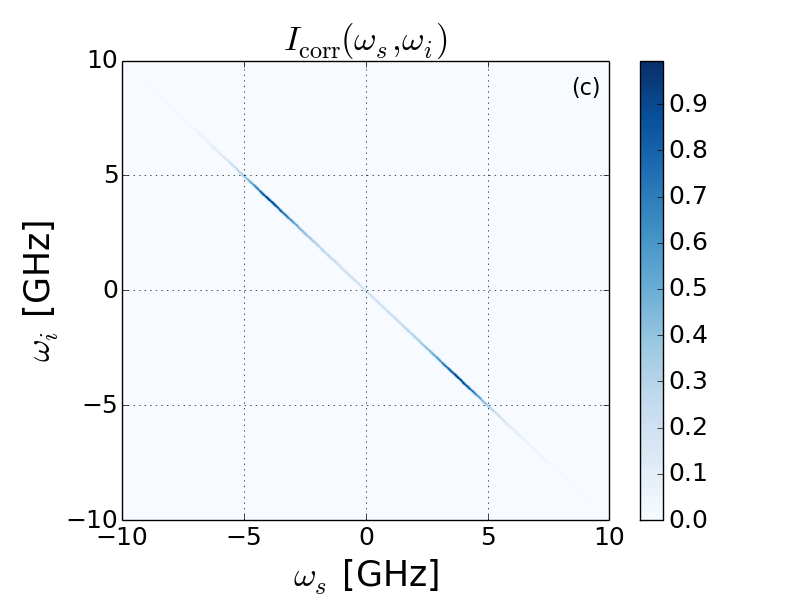}
\caption{(Colour online) Correlated part $I_\mathrm{corr}$ of joint spectral intensity distribution, scaled to unit maximum, of signal and idler photon pairs for a pump with (a) $\Delta_P=0, N_P=10$, (b) $\Delta_P=3\barGamma, N_P=10$, and (c) $\Delta_P=0, N_P=2\times 10^8$. Ring parameters are taken as $\barGamma=1$ GHz and $\Lambda=10$ Hz. The splitting evident in (b) arises from the pump detuning, whereas in (c) the XPM-induced detuning of the signal and idler ring modes is responsible.}\label{fig::JSI_splitting_plots}
\end{figure}

For low power cw pumps, wherein multi-pair generation is insignificant, the uncorrelated part of the JSI $I_\mathrm{uncorr}$ is negligible and $I_\mathrm{corr}$ dominates. The JSI then takes the form of a narrow antidiagonal line corresponding to the energy-conserving condition $\omega_s+\omega_i-2\Delta_P=0$. For $\Delta_P=0$, the line is singly peaked, as illustrated in Fig. \ref{fig::JSI_splitting_plots}(a). For nonzero $\Delta_P$ the line is distributed among two peaks separated by $2\Delta_P$, as evident in Fig. \ref{fig::JSI_splitting_plots}(b), consistent with the single photon spectrum derived in the previous section. At higher powers, such as in Fig. \ref{fig::JSI_splitting_plots}(c), this splitting can also arise from XPM-induced signal and idler detuning even for a pump with $\Delta_P=0$.  When the splitting is due to XPM-induced signal and idler detuning, the JSI remains centred on the unperturbed ring resonances. On the other hand, when pump detuning is responsible for the splitting, the JSI is translated by $\Delta_P$ along both frequency axes. 

\begin{figure}
\includegraphics[width=1.0\columnwidth]{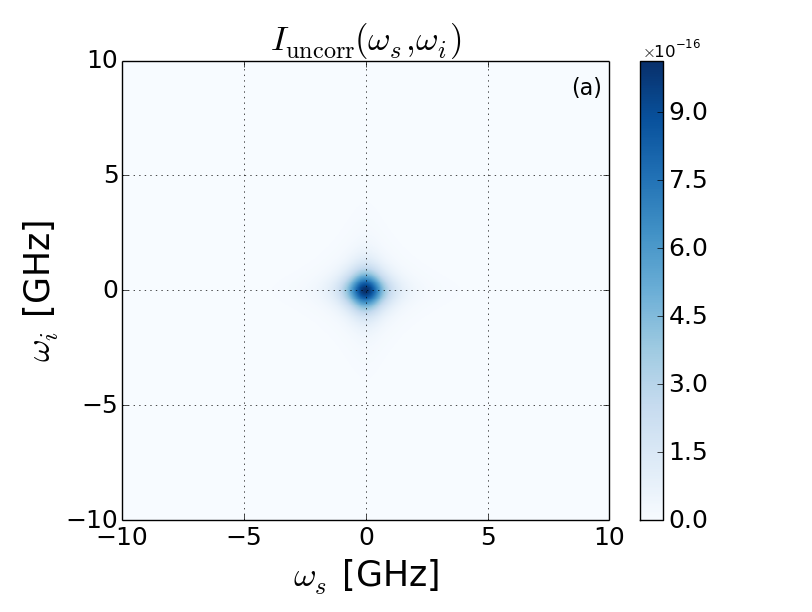}
\includegraphics[width=1.0\columnwidth]{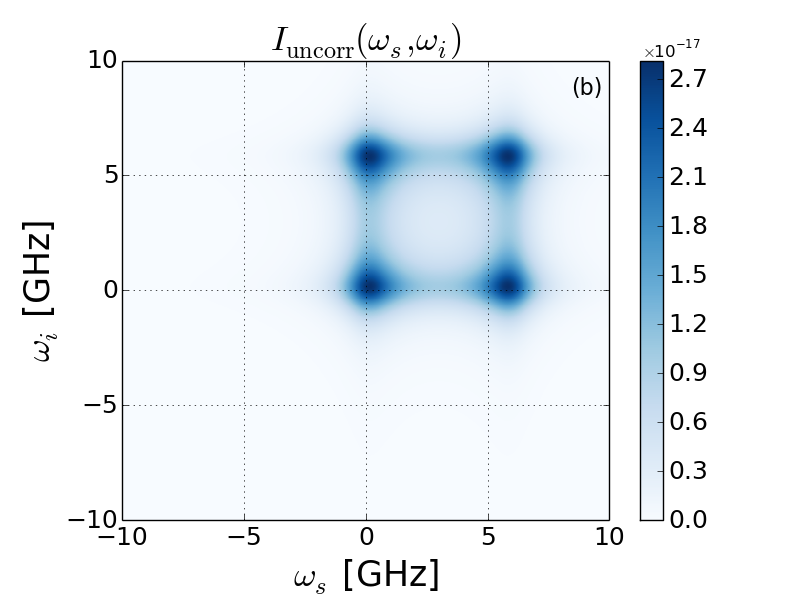}
\includegraphics[width=1.0\columnwidth]{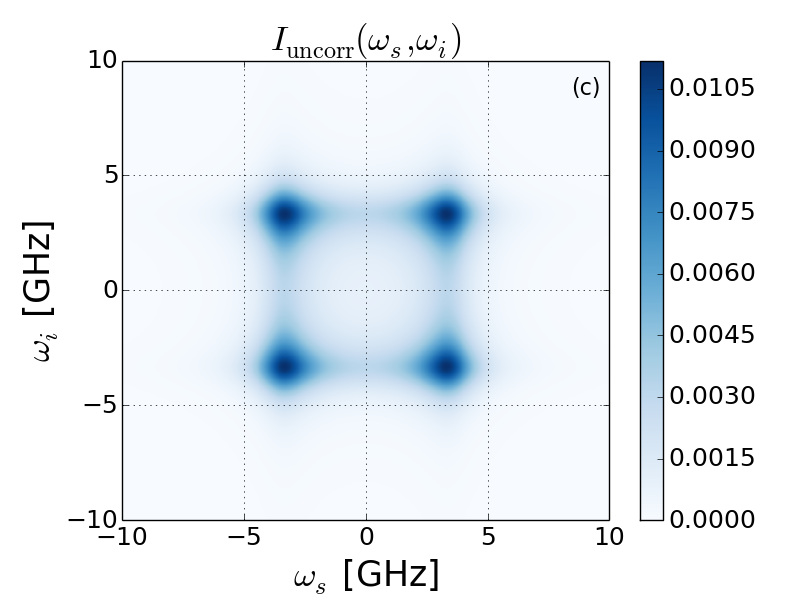}
\caption{(Colour online) Uncorrelated part $I_\mathrm{uncorr}$ of  joint spectral intensity distribution, scaled to unit maximum, of signal and idler photon pairs. Pump parameters are (a) $\Delta_P=0, N_P=10$, (b) $\Delta_P=3\barGamma, N_P=10$, and (c) $\Delta_P=0, N_P=2\times 10^8$. Ring parameters are taken as $\barGamma=1$ GHz and $\Lambda=10$ Hz.}\label{fig::JSI_uncorrelated_piece}
\end{figure}

In Fig. \ref{fig::JSI_uncorrelated_piece} the uncorrelated contribution $I_\mathrm{uncorr}$ to the JSI is plotted for the same pump parameters as in Fig. \ref{fig::JSI_splitting_plots}. The weight of the uncorrelated contribution is extremely small compared to the correlated contribution at low powers, as indicated by the scales in Figs. \ref{fig::JSI_splitting_plots} and \ref{fig::JSI_uncorrelated_piece}, but grows to an appreciable level at high powers. For $\Delta_P=0$, as in Fig. \ref{fig::JSI_uncorrelated_piece}(a), at low $N_P$ the uncorrelated part of the JSI displays a single peak centred at the origin.  In the regimes that give rise to split lineshapes, as illustrated in Fig. \ref{fig::JSI_uncorrelated_piece}(b) and \ref{fig::JSI_uncorrelated_piece}(c), the uncorrelated contribution takes the form of four distinct peaks, symmetrically placed about the centre of the overall distribution. Two of these peaks lie on the antidiagonal, overlapping with the correlated contribution. The remaining two lie on the diagonal, and would therefore appear to violate energy conservation if assumed to correspond to signal and idler photons that originated from the same pair. It is therefore natural to interpret these peaks as corresponding to signal and idler photons that are detected from \emph{separate} pairs. As these uncorrelated, ``non-energy conserving" peaks are well separated from the correlated part of the JSI, they are uncontaminated by the correlated contribution to the JSI. The properties of photon pairs detected in these peaks would therefore be expected to differ from those detected in the antidiagonal peaks. We intend to investigate such properties in future work.

\begin{figure}
\includegraphics[width=1.0\columnwidth]{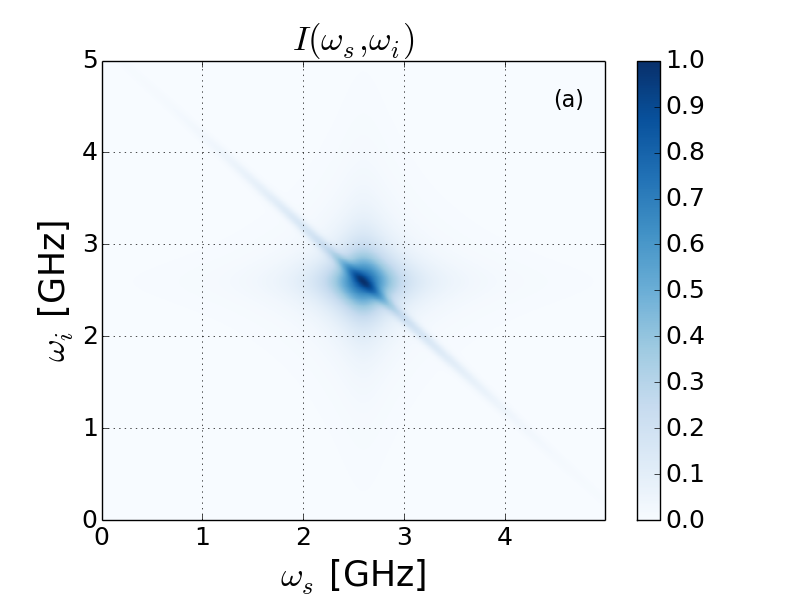}
\includegraphics[width=1.0\columnwidth]{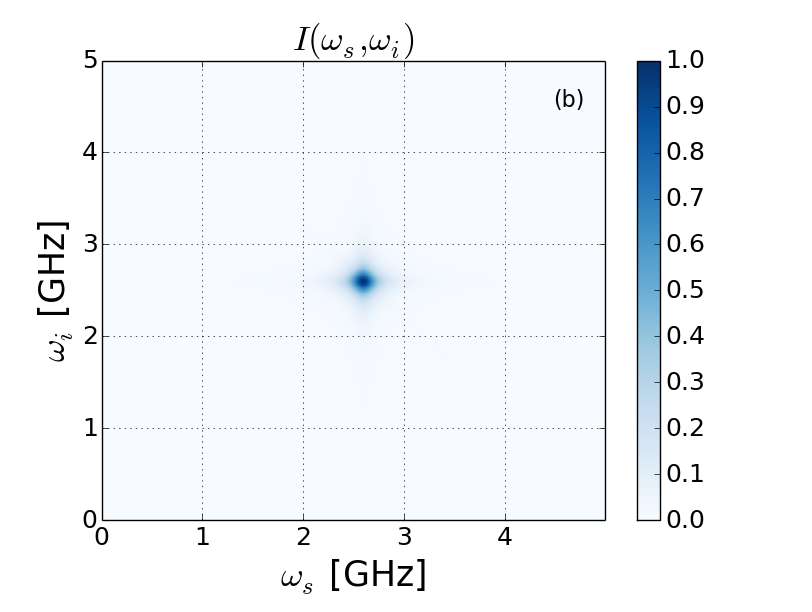}
\caption{(Colour online) Joint spectral intensity distribution, scaled to unit maximum, of signal and idler photon pairs for (a) $\Delta_P=1.5\Delta_\mathrm{critical}$ with $N_P=9.8\times 10^7$ (90\% of OPO threshold) and (b) $\Delta_P=1.5\Delta_\mathrm{critical}$ with $N_P=1\times 10^8$ (95\% of OPO threshold). Ring parameters are taken as $\barGamma=1$ GHz and $\Lambda=10$ Hz.}\label{fig::JSI_real_rhobar}
\end{figure}

The form of the JSI depends qualitatively on whether $\rhobar$ is imaginary or real, a behaviour we saw earlier in the single photon spectrum. When $\rhobar$ is imaginary, and thus contributes to the frequency terms in the denominators of Eqs. () and (), a splitting in the JSI appears. When $\rhobar$ is real, and acts as an effective correction to the linewidth $\barGamma$, the JSI is localized to a single line arising from the overlap of $I_\mathrm{corr}$ with a single peak in $I_\mathrm{uncorr}$. For sufficiently detuned pumps, in the regime of real $\rhobar$ the uncorrelated contribution can be large enough to be visible on the JSI plot without exaggeration or scaling. Indeed, for supercritical detunings $|\Delta_P|>\Delta_\mathrm{critical}$, as the OPO threshold is approached and $\rhobar\to\barGamma$ the uncorrelated contribution vastly dominates over the correlated contribution, as seen in Fig. \ref{fig::JSI_real_rhobar}. This is an expected consequence of the rapid growth in the photon pair generation rate in this regime -- multiple photon pairs are generated in sufficiently large quantities that joint detection of a signal and idler photon originating from the same energy-conserving pair is unlikely relative to the probability of detecting a signal and idler photon which originated from separate pairs and thus obey no relationship in energy. Another effect seen as $\rhobar\to\barGamma$ is the narrowing of the entire JSI distribution to a small point-like peak centred on $(\omega_s,\omega_i)=(\Delta_P,\Delta_P)$. Within our  idealization of a zero-bandwidth cw pump the area of this point would be limited only by the frequency resolution of the JSI measurement scheme, though in actual experiments the finite pump bandwidth would serve as a fundamental lower bound for the overall extent of the JSI.

Perhaps the most definitive experimental indication of strongly driven effects lies in the top-right and bottom-left uncorrelated peaks of the JSI distribution for a detuned pump as in Fig. \ref{fig::JSI_uncorrelated_piece}(b). For sufficient detunings these peaks are well separated from the antidiagonal and thus easily distinguished from the correlated part of the JSI. For low powers, wherein only one photon pair is generated in the ring at any given time, they would be entirely absent from the measured JSI. As the power increases, \emph{any} non-spurious coincidence detection of photons in these regions indicates multi-pair generation, as photons generated in those peaks do not conserve energy and therefore must be associated with separate, independently produced pairs. 
\section{Conclusion}
We have investigated the strongly driven regime of spontaneous four-wave mixing in microring resonators for a cw pump input. A nonperturbative, exact analytic solution to the semiclassical equations of motion within the undepleted pump approximation was developed, which permits the calculation of any physical quantity related to the outgoing signal and idler fields while fully taking into account intraring scattering losses. The effects of self- and cross- phase modulation, as well as multi-pair generation, were found to drastically alter the nature of the photon pair generation process at high powers. A critical pump detuning of $\Delta_\mathrm{critical}=\sqrt{3}\;\barGamma$, where $\barGamma$ is the total effective linewidth of the ring resonances, was found to divide the behaviour of the system into two regimes. For supercritically detuned pumps, a region of optical bistability of the pump mode is predicted, and a threshold emerges for optical parametric oscillation of the signal and idler modes. Pump power-dependent splitting of the generated signal and idler photon spectra was uncovered, arising from both pump detuning and cross-phase modulation. In certain intermediate-power regimes, dramatic narrowing of the spectral linewidth of generated signal and idler photons associated with the approach to optical parametric oscillation was found. The joint spectral intensity distribution (JSI) was analysed, and found to consist of separate uncorrelated and correlated contributions. The correlated contribution is negligible at low powers, but becomes significant as multi-pair generation becomes appreciable at higher powers. In the regime of spectral splitting, the uncorrelated part of the JSI displays an intriguing quadruplet of peaks, two of which are well separated from the correlated part. An optimal detuning strategy was derived in which the pump detuning is chosen to exactly cancel the effect of self-phase modulation at each input power,  maximizing the intraring pump intensity. By detuning the pump in this manner the effects of both spectral splitting and bandwidth reduction are eliminated, and the photon pair generation rate continues to scale quadratically with the pump input even for arbitrarily high powers.

Three simple experimental tests of our predictions in the strongly driven regime were proposed:
\begin{enumerate}
\item For fixed subcritical nonzero detunings the photon pair generation rate as a function of input pump power is predicted to have a local maximum at intermediate powers, followed by a decreasing approach to an asymptotic level at high powers.
\item The single photon spectra of the outgoing signal and idler fields are predicted to show spectral splitting proportional to the pump detuning at low powers, followed by a regime of a singly peaked spectrum with pump power-dependent narrowing of bandwidth at intermediate powers, and finally resuming a doublet structure at high powers.
\item The presence of two non-energy conserving peaks lying on the diagonal of the joint spectral intensity distribution, which are a consequence of multi-pair generation, is predicted to occur for sufficiently large pump powers.
\end{enumerate}
Our analysis was restricted to cw pump inputs; studying how these strongly driven phenomena are altered for short pulses requires a numerical approach. Additionally, a slightly more sophisticated solution is required to fully study the regime of optical parametric oscillation, in which the undepleted pump approximation breaks down. We intend to extend our techniques to treat these regimes future publications.

\appendix
\section{Calculation of nonlinear coupling constants}\label{appendix:lambda}
To estimate the nonlinear coupling constants $\Lambda$, $\eta$ and $\zeta$, we present in this section a derivation of the nonlinear sector $H_\mathrm{NL}$ of the ring Hamiltonian. For the moment we imagine the ring has been decoupled from both the physical and phantom channels, so that it is idealized as a perfect, isolated cavity. We expand the electric field $\mathbf{E}(\mathbf{r})$ and electric displacement field $\mathbf{D}(\mathbf{r})$ in the ring in terms of discrete ring modes $\mathbf{E}_\alpha(\mathbf{r})$ and $\mathbf{D}_\alpha(\mathbf{r})$ as
\begin{eqnarray}\label{modes_defn}
\mathbf{E}(\mathbf{r})&=&\sum_\alpha \sqrt{\frac{\hbar\omega_\alpha}{2}}b_\alpha \mathbf{E}_\alpha(\mathbf{r}) + \mathrm{H.c.}, \nonumber \\
\mathbf{D}(\mathbf{r})&=&\sum_\alpha \sqrt{\frac{\hbar\omega_\alpha}{2}}b_\alpha \mathbf{D}_\alpha(\mathbf{r}) + \mathrm{H.c.},
\end{eqnarray}
where $\omega_\alpha$ are the mode frequencies and $b_\alpha$ the associated annihilation operators. The contribution to the ring Hamiltonian arising from the third-order nonlinear susceptibility can be written \cite{Sipe2004} as
\begin{eqnarray}
\lefteqn{H_\mathrm{NL} =}\\
& & -\frac{1}{4\epsilon_0}\int d\mathbf{r} \Gamma^{ijkl}_{(3)}(\mathbf{r})D^i(\mathbf{r})D^j(\mathbf{r})D^k(\mathbf{r})D^l(\mathbf{r}) \nonumber
\end{eqnarray}
with implied summation over repeated lowercase Roman indices, where $\epsilon_0$ is the permittivity of vacuum and $\Gamma^{ijkl}_{(3)}(\mathbf{r})$ represents the nonlinear response coefficients. Within the rotating wave approximation, only keeping relevant terms for the pump, signal and idler modes, we obtain
\begin{eqnarray}
\lefteqn{H_\mathrm{NL}=} \\
&-&\frac{1}{4\epsilon_0}\left(\frac{4!}{2!1!1!}\right)\frac{\hbar\omega_P}{2}\sqrt{\frac{\hbar\omega_S}{2}\frac{\hbar\omega_I}{2}} Q_{SIPP} b_S^\dagger b_I^\dagger b_P b_P \nonumber \\
&-&\frac{1}{4\epsilon_0}\left(\frac{4!}{2!1!1!}\right)\frac{\hbar\omega_P}{2}\sqrt{\frac{\hbar\omega_S}{2}\frac{\hbar\omega_I}{2}} Q_{PPIS} b_P^\dagger b_P^\dagger b_I b_S \nonumber \\
&-&\frac{1}{4\epsilon_0}\left(\frac{4!}{2!2!}\right)\left(\frac{\hbar\omega_P}{2}\right)^2 Q_{PPPP} b_P^\dagger b_P^\dagger b_P b_P \nonumber \\
&-&\frac{1}{4\epsilon_0}\left(\frac{4!}{1!1!1!1!}\right)\left(\frac{\hbar\omega_P}{2}\frac{\hbar\omega_S}{2}\right) Q_{SPSP} b_S^\dagger b_P^\dagger b_S b_P \nonumber \\
&-&\frac{1}{4\epsilon_0}\left(\frac{4!}{1!1!1!1!}\right)\left(\frac{\hbar\omega_P}{2}\frac{\hbar\omega_I}{2}\right) Q_{IPIP} b_I^\dagger b_P^\dagger b_I b_P \nonumber,
\end{eqnarray}
where the constants $Q_{IJKL}$ are given by
\begin{eqnarray}
\lefteqn{Q_{IJKL} = } \\
& &\int d\mathbf{r}\left(\Gamma^{ijkl}_{(3)}(\mathbf{r}) (D_I^i(\mathbf{r}))^*(D_J^j(\mathbf{r}))^*D_K^k(\mathbf{r})D_L^l(\mathbf{r})\right)\nonumber.
\end{eqnarray}
 As is typically done for dispersive media we take \cite{Bhat2006}
\begin{eqnarray}
\lefteqn{\Gamma^{ijkl}_{(3)}(\mathbf{r})} \\
&=&\frac{\chi_{(3)}^{ijkl}(\mathbf{r})}{\epsilon_0^2 n^2(\mathbf{r};\omega_1)n^2(\mathbf{r};\omega_2)n^2(\mathbf{r};\omega_3)n^2(\mathbf{r};\omega_4)},\nonumber
\end{eqnarray}
where $n(\mathbf{r};\omega)$ is the linear refractive index of the ring medium at frequency $\omega$, and $\chi_{(3)}^{ijkl}(\mathbf{r})$ is the frequency-dependent nonlinear susceptibility. To evaluate the coefficients $Q_{IJKL}$ we introduce co-ordinates for the ring $\mathbf{r}_\perp$ and $l_\phi$, such that the volume element
\begin{eqnarray}
d\mathbf{r} = \rho d\rho d\phi dz
\end{eqnarray} 
can be written as
\begin{eqnarray}
d\mathbf{r} = \frac{\rho d\rho dz}{R}dl_\phi = d\mathbf{r}_\perp dl_\phi,
\end{eqnarray}
where $R$ is the nominal ring radius and $l_\phi=R\phi$, which varies from $0$ to $2\pi R \equiv L$, the nominal ring circumference. The co-ordinate $\mathbf{r}_\perp$ is understood as shorthand for the pair $(\rho,z)$. Writing the mode fields $\mathbf{E}_\alpha(\mathbf{r})$ as
\begin{eqnarray}
\mathbf{E}_\alpha(\mathbf{r}) = \frac{\mathbf{e}_\alpha(\mathbf{r}_\perp)e^{i k_\alpha l_\phi}}{\sqrt{L}},
\end{eqnarray}
where $k_\alpha=2\pi n_\alpha/L$ for integer $n_\alpha$, we can simplify $H_\mathrm{NL}$ to
\begin{eqnarray}\label{H_expansion}
\lefteqn{H_\mathrm{NL} =} \\
&-& \frac{3}{L^2}\frac{\hbar\omega_P}{2}\sqrt{\frac{\hbar\omega_S}{2}\frac{\hbar\omega_I}{2}} Q_{SIPP}'b_S^\dagger b_I^\dagger b_P b_P \nonumber \\
&-& \frac{3}{L^2}\frac{\hbar\omega_P}{2}\sqrt{\frac{\hbar\omega_S}{2}\frac{\hbar\omega_I}{2}} Q_{PPIS}'b_P^\dagger b_P^\dagger b_S b_S
 \nonumber \\
&-& \frac{3}{2L^2}\left(\frac{\hbar\omega_P}{2}\right)^2 Q_{PPPP}'b_P^\dagger b_P^\dagger b_P b_P \nonumber \\
&-& \frac{6}{L^2}\left(\frac{\hbar\omega_P}{2}\frac{\hbar\omega_S}{2}\right)^2 Q_{SPSP}'b_S^\dagger b_P^\dagger b_S b_P \nonumber \\
&-& \frac{6}{L^2}\left(\frac{\hbar\omega_P}{2}\frac{\hbar\omega_I}{2}\right)^2 Q_{IPIP}'b_I^\dagger b_P^\dagger b_I b_P \nonumber,
\end{eqnarray}
in which the reduced constants $Q_{IJKL}'$ are given by
\begin{eqnarray}\label{Q_reduced}
\lefteqn{Q_{IJKL}' = \frac{1}{\sqrt{Z_I Z_J Z_K Z_L}}} \\
&\times& \bigg[\int d\mathbf{r}_\perp d l_\phi \epsilon_0\chi_{(3)}^{ijkl}(\mathbf{r}_\perp,l_\phi)(e^i_I(\mathbf{r}_\perp))^*(e^j_J(\mathbf{r}_\perp))^* \nonumber \\
& &\;\;\;\;\times
e^k_K(\mathbf{r}_\perp)e^l_L(\mathbf{r}_\perp)e^{i(k_K + k_L - k_I - k_J) l_\phi}\bigg].
\end{eqnarray}
The modes (\ref{modes_defn}) are normalized \cite{Bhat2006} such that 
\begin{eqnarray}
\lefteqn{Z_\alpha =} \\
& & \frac{1}{L}\int d\mathbf{r}_\perp dl_\phi \epsilon_0 n^2(\mathbf{r}_\perp;\omega_\alpha)\mathbf{e}^*_\alpha(\mathbf{r}_\perp)\cdot \mathbf{e}_\alpha(\mathbf{r}_\perp)\gamma_\mathrm{gp}(\mathbf{r}_\perp;\omega_\alpha),\nonumber \\
&=& 1,
\end{eqnarray}
where $\gamma_\mathrm{gp}(\mathbf{r}_\perp ;\omega_\alpha)$ is the ratio of the group and phase velocities of the ring medium at each frequency and spatial point. However, we display the $Z_\alpha$ explicitly in (\ref{Q_reduced}) so that the expression can be used regardless of whether or not the $\mathbf{e}_\alpha(\mathbf{r}_\perp)$ are normalized such that $Z_\alpha=1$. 

To estimate the constants $Q'_{IJKL}$ we approximate the ratio $\gamma_{gp}\approx 1$ everywhere, and assume the $\mathbf{e}_\alpha(\mathbf{r}_\perp)$ to be of uniform magnitude within the ring and vanish elsewhere. We take this uniform magnitude to be unity and assume that for the modes of interest $(2k_P - k_I - k_S)L\ll 1$. For modes with polarization mainly perpendicular to the ring plane, the relevant susceptibility will be $\chi_{(3)}^{zzzz}\equiv \chi_{(3)}$, independent of position in the ring; this can be immediately generalized to treat other mode polarizations. We then have
\begin{eqnarray}
Z_\alpha = \epsilon_0 n^2 A,
\end{eqnarray}
where A is the cross-sectional area of the ring. Taking $\omega_S\approx\omega_I\approx\omega_P$ in the prefactors of (\ref{H_expansion}), we finally obtain
\begin{eqnarray}
H_{\mathrm{NL}} &\approx& \left(\hbar\Lambda b_P b_P b_S^\dagger b_I^\dagger + \mathrm{H.c.}\right) + \hbar\eta b_P^\dagger b_P^\dagger b_P b_P \nonumber \\
&+& \hbar\zeta\left(b_S^\dagger b_P^\dagger b_S b_P + b_I^\dagger b_P^\dagger b_I b_P\right),
\end{eqnarray}
where
\begin{eqnarray}
\Lambda = \frac{3\hbar\omega_P^2\chi_{(3)}}{4\epsilon_0 n^4 LA}
\end{eqnarray}
and $\eta=\Lambda/2$, $\zeta=2\Lambda$. In terms of the more experimentally accessible nonlinear refractive index $n_2=3\chi_{(3)}/4\epsilon_0cn^2$ this becomes
\begin{eqnarray}
\Lambda = \frac{\hbar\omega_P^2 c n_2}{n^2 L A},
\end{eqnarray}
which is in line with the results of similar derivations \cite{Andersen2015}.

\section{An operational definition of the joint spectral intensity}\label{appendix:JSI}
When characterizing a source of entangled photon pairs, the \emph{joint spectral intensity} (JSI) distribution is often introduced in the low power limit, when the state of the signal and idler modes is well approximated by 
\begin{eqnarray}\label{two_photon_JSI}
|\psi_{SI}\rangle &=& p_\mathrm{vac}|\mathrm{vac}\rangle \\
&+& \int d\omega_s\int d\omega_i f(\omega_s,\omega_i)a_S^\dagger(\omega_s) a_I^\dagger(\omega_i)|\mathrm{vac}\rangle,\nonumber
\end{eqnarray}
where $|p_\mathrm{vac}|^2<1$ is a constant and the $a_J^\dagger(\omega)$ refer to the creation operators at frequency $\omega$ for the signal and idler modes\cite{Grice2001}. The unsymmetrized and unnormalized JSI for the such a state is defined as $|f(\omega_s,\omega_i)|^2$, and is proportional to the probability density per unit time of jointly detecting a signal and idler photon pair with respective frequencies $\omega_s$ and $\omega_i$. For strongly pumped sources, when multiple photon pairs are generated in significant quantities  so that higher order terms involving more than two creation operators appear in the state, it is less straightforward to define a single function that characterizes the energy relationship between simultaneously detected signal and idler photons. Instead, one can operationally extend the definition of the JSI to strongly pumped sources by calculating for arbitrary input power the outcome of experiments designed to measure $|f(\omega_s,\omega_i)|^2$ in the low power limit. In this section we develop such a calculation for a typical measurement scheme employed to measure the JSI of photon pairs produced in a microring resonator.
\begin{figure}
\includegraphics[width=1.0\columnwidth]{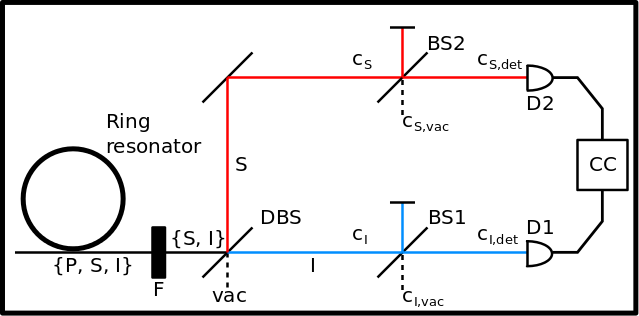}
\caption{(Colour online) Schematic of experimental setup for measuring the joint spectral intensity distribution. Pump, signal and idler $\{P,S,I\}$ outputs from the ring resonator are incident on a filter $F$ that removes the pump component from the beam. Signal and idler fields with modes $c_S$ and $c_I$ are then separated by dichroic beamsplitter DBS, and independently filtered by monochromators, which are implemented by frequency-dependent beamsplitters BS1 and BS2. Each monochromator-beamsplitter transmits in a small window $\delta\omega_\mathrm{trans}$ about $\omega_i$ and $\omega_s$, respectively. Broadband photodetectors D1 and D2 measure the detector modes $c_{S,\mathrm{det}}$ and $c_{I,\mathrm{det}}$, and are connected to coincidence counter CC to register joint detection events within a temporal resolution of $\delta t$. Vacuum is input to the empty ports of BS1, BS2 and DBS.}\label{fig:JSI_schematic}
\end{figure}

We consider a standard experimental setup \cite{Kim2005} to measure coincidence rates between signal and idler photons of particular frequencies as illustrated in Fig. \ref{fig:JSI_schematic}. The signal and idler fields are separated, and each field is sent through a separate monochromator set to transmit photons in some small range $\delta\omega_\mathrm{trans}$ about a centre frequency $\omega_s$ for the signal and $\omega_i$ for the idler. Placed after each monochromator are broadband photodetectors connected to a coincidence counter to identify simultaneously detected signal and idler photons. The transmission frequencies $\omega_s$ and $\omega_i$ are independently controllable, and correspond to a single point (or, more accurately, single bin) on the JSI plot, which is produced by scanning through $\omega_s$ and $\omega_i$ and measuring the corresponding coincidence rate. The transmission width is chosen to be much smaller than the linewidth of the measured photons, $\delta\omega_\mathrm{trans}\ll\barGamma$, so that the full 2D spectrum can be resolved.

The monochromators can be simply modelled as frequency-dependent beamsplitters. Provided both the signal and idler arms of the experiment are balanced, the spatial dependence of the fields after the ring can be suppressed; all fields in this section are understood to be evaluated immediately after the ring-channel coupling point. We introduce annihilation operators $c_S(\omega_s)$ and $c_I(\omega_i)$ for the ring output fields $\barpsi_{S>}(t)$ and $\barpsi_{I>}(t)$ as in Eq. (\ref{amplitude_defn}). We can then apply the appropriate transformation to obtain the annihilation operators $c_{J,\mathrm{det}}(\omega_j)$ for the fields seen by the detectors placed after the monochromators. For the signal, we have
\begin{eqnarray}
c_{S,\mathrm{det}}(\omega) = \hat{T}(\omega-\omega_s)c_S(\omega) + \hat{R}(\omega-\omega_s)c_{S,\mathrm{vac}}(\omega),\;\;\;\;\;
\end{eqnarray}
in which $c_{S,\mathrm{vac}}$ refer to the modes on the other input port of the monochromator-beamsplitter, into which only vacuum is present. The transmission and reflection functions $\hat{T}(\omega)$ and $\hat{R}(\omega)$ determine which frequencies are transmitted by the the monochromator. For example, a simple filter may be modelled by a transmission function with a rectangular frequency profile,
\begin{eqnarray}
\hat{T}(\omega) = \begin{cases}
1, & -\frac{\delta\omega_\mathrm{trans}}{2} < \omega < \frac{\delta\omega_\mathrm{trans}}{2} \\
0, & \mathrm{otherwise}. \end{cases}
\end{eqnarray}
The exact choice of $\hat{T}(\omega)$ is not important for our purposes; for simplicity we only assume $\hat{T}(\omega)$ is a real, sufficiently narrow, symmetric function of $\omega$. The reflection function will be irrelevant, though it will satisfy the usual restrictions to correctly model a beamsplitter. 

In exactly the same manner, modes $c_{I,\mathrm{det}}(\omega)$ seen by the detectors of the idler arm can be introduced. We can then write down the fields measured by each detector in the usual way,
\begin{eqnarray}
\barpsi_{J,\mathrm{det}}(t) = \int\frac{d\omega_j}{\sqrt{2\pi}}c_{J,\mathrm{det}}(\omega_j)e^{-i\omega_j t}.
\end{eqnarray}
In a typical coincidence measurement the signal detector is continuously activated, and detection of a signal photon at time $t$ is used to trigger the activation of the idler detector (which is placed at a small delay relative to the signal detector) for a very short time $\delta t$, so that the idler detector samples the idler field during the time interval $[t-\delta t/2,t+\delta t/2]$. The average rate $I(\omega_s,\omega_i; t)$ at time $t$ of coincident detection events at $\omega_s$ and $\omega_i$ of the signal and idler detectors is given by the standard Glauber formula involving the fields at each detector \cite{GerryKnight2004},
\begin{eqnarray}
\lefteqn{I(\omega_s,\omega_i;t) =}\\
& &\lim_{T\to\infty}\Bigg[\frac{1}{T} \int\displaylimits_{t-T/2}^{t+T/2}dt'\int\displaylimits_{t'-\delta t/2}^{t'+\delta t/2} dt'' \nonumber \\
& &v^2 \langle \barpsi_{S,\mathrm{det}}^\dagger(t')\barpsi_{I,\mathrm{det}}^\dagger(t'')\barpsi_{S,\mathrm{det}}(t')\barpsi_{I,\mathrm{det}}(t'') \rangle\Bigg].
\end{eqnarray}
In steady state the expectation value depends only on time difference $|t''-t'|$, which in the integrand is at most $\delta t$. Provided the coincidence resolution time $\delta t$ is much smaller than the timescale on which the expectation value varies (in our case $\barGamma^{-1}$), this expression for $I(\omega_s,\omega_i;t)$ is then well approximated by
\begin{eqnarray}
\lefteqn{I(\omega_s,\omega_i;t) \approx} \\
& & v^2 \delta t\langle \barpsi_{S,\mathrm{det}}^\dagger(t)\barpsi_{I,\mathrm{det}}^\dagger(t)\barpsi_{S,\mathrm{det}}(t)\barpsi_{I,\mathrm{det}}(t) \rangle.\nonumber
\end{eqnarray}

Proceeding to expand the detector fields in terms of their constituent modes, we find
\begin{eqnarray}
\lefteqn{I(\omega_s,\omega_i;t) =} \\
& &\frac{v^2\delta t}{(2\pi)^2}\int d\nu_1 ...\int d\nu_4\bigg[e^{i(\nu_1+\nu_2-\nu_3-\nu_4)} \nonumber \\ &\times&\hat{T}^*(\nu_1-\omega_s)\hat{T}^*(\nu_2-\omega_i) \hat{T}(\nu_3-\omega_s)\hat{T}(\nu_4-\omega_i)\nonumber \\
&\times&\langle c_S^\dagger(\nu_1)c_I^\dagger(\nu_2)c_S(\nu_3)c_I(\nu_4)\rangle\bigg].\nonumber
\end{eqnarray}
By writing the operators $c_J(\nu_i)$ in terms of their respective parent fields and then carrying out the integration over each $\nu_i$, we arrive at Eq. (\ref{JSI_definition}), where $T(t)$ is the Fourier transform of the transmission function $\hat{T}(\omega)$; see Eq. (\ref{T_defn}).

In obtaining (\ref{JSI_definition}) we have again used the fact that in steady state the expectation value is invariant with respect to time translations by $t$ in each argument. The expression (\ref{JSI_definition}) is manifestly real, independent of time, and in the limit of small $\delta\omega_\mathrm{trans}$ indeed reproduces the single-pair JSI $|f(\omega_s,\omega_i)|^2$ when calculated with an initial state of the form (\ref{two_photon_JSI}). 

\begin{acknowledgments}
This work was financially supported by the Natural Sciences and Engineering Research Council of Canada.
\end{acknowledgments}

\bibliography{CW_draft}

\end{document}